# Geometrical portrait of Multipath error propagation in GNSS Direct Position Estimation

Jihong Huang, Rong Yang*, Member, IEEE, Wei Gao, Xingqun Zhan, Senior Member, IEEE, Zheng Yao, Senior Member, IEEE

Abstract—Direct Position Estimation (DPE) is a method that directly estimate position, velocity, and time (PVT) information from cross ambiguity function (CAF) of the GNSS signals, significantly enhancing receiver robustness in urban environments. However, there is still a lack of theoretical characterization on multipath errors in the context of DPE theory. Geometric observations highlight the unique characteristics of DPE errors stemming from multipath and thermal noise as estimation bias and variance respectively. Expanding upon the theoretical framework of DPE noise variance through geometric analysis, this paper focuses on a geometric representation of multipath errors by quantifying the deviations in CAF and PVT solutions caused by off-centering bias relative to the azimuth and elevation angles. A satellite circular multipath bias (SCMB) model is introduced, amalgamating CAF and PVT errors from multiple satellite channels. The boundaries for maximum or minimum PVT bias are established through discussions encompassing various multipath conditions. The correctness of the multipath geometrical portrait is confirmed through both Monte Carlo simulations and urban canyon tests. The findings indicate that the maximum PVT bias depends on the largest multipath errors observed across various satellite channels. Additionally, the PVT bias increases with satellite elevation angles, influenced by the CAF multipath bias projection. This serves as a reference for selecting DPE satellites from a geometric standpoint, underscoring the importance of choosing a balanced combination of high and low elevation angles to achieve an optimal satellite geometry configuration.

## I. Introduction

As the demand for positioning, navigation, and timing (PNT) increases, the requirements for the reliance and robustness of Global Navigation Satellite System (GNSS) receivers are also growing. Direct Position Estimation (DPE) is a technique that differs from traditional synchronization-parameter-based positioning approach by directly estimating position, velocity, and timing (PVT) from cross ambiguity function (CAF) of the GNSS signals, theoretically and experimentally improving receiver robustness and reliability [1], [2]. This capability is particularly advantageous in chanllenging environment where DPE does not necessitate constant signal tracking. It enables the incorporation of all visible satellites in positioning via CAF accumulation, extending beyond the confines of conventional tracking boundaries. However, with ongoing research, there is room for improvement in the performance of current DPE algorithms in dense multipath environments like urban canyons. Experimental results indicate that DPE PVT solution may diverge in the presence of serve multipath interference [3]. The multipath delay can deform CAF, and significantly increase PVT estimation errors [4].

To tackle multipath issues in DPE, some explorations have been made [5]–[10]. For example, reference [5] applied the robust interference mitigation (RIM) approach to enhance the CAF in DPE framework. Reference [6] used 3D building models for multipath inversion, leveraging non-line-of-sight (NLOS) signals to assist in positioning. Reference [7] utilized environmental information and specular reflection theory to constrain the candidate search range in DPE receiver. Reference [8] placed the azimuth constraints to approximate the theoretical maximum likelihood estimate (MLE) of DPE. Reference [9] reduced the multipath impact on DPE performance by removing interfered pseudoranges. And in our previous work [10], we attempted to mitigate multipath error using multifrequency combination in DPE receiver design to enhance its accuracy and robustness.

Nonetheless, there remains a deficiency in theoretical characterization on multipath effects within the DPE framework, rendering optimization enhancement challenging in urban environment. Due to the coupling effects of noise and multipath on the GNSS signals, it is extremely challenging to identify and quantify the components of multipath error from CAF analytically. By simple approximation of the multipath error as a part of additive noise in signal model can introduce severe divergence in CAF and lead to incorrect PVT estimations [11]. For improvement, a probability distribution function (PDF) for multipath error was built and the DPE estimator was redesigned based on MLE criteria [12]. However, there remains a deficiency in analytical derivations concerning DPE errors arising from noise

J. Huang, R. Yang, W. Gao and X. Zhan are with School of Aeronautics and Astronautics, Shanghai Jiao Tong University, Shanghai, China. E-mail:(jihong.huang@sjtu.edu.cn, rongyang@sjtu.edu.cn, gaowei1515@sjtu.edu.cn, xqzhan@sjtu.edu.cn). R. Yang is the correspondance author.

and multipath components. Specifically, the DPE error transfer or projection process from correlation domain to navigation domain, or equivalently from CAF to PVT solutions, has not been precisely characterized.

Physically, multipath errors arise from signal reflections, influenced by the spatial structure location and satellite distribution, exhibiting distinct geometry patterns and predictability. In contrast, noise is stochastic, characterized by a random distribution. Thus, errors induced by multipath and noise fundamentally differ in nature. However, theoretical distinguishing between noise and multipath from a signal estimation perspective remains challenging due to the simplified mathematical derivation under the ideal Gaussian linearity hypothesis. By revisiting the DPE error transfer process from a geometric viewpoint in our previous work [13], the noise variance boundary has been meticulously crafted by leveraging real data processing outcomes, accounting for the projection from CAF to PVT across a range of satellite elevation and azimuth angles.

Upon this theoretical geometric foundation, this paper mainly focuses on geometric representation on DPE multipath errors, by quantifying the multipath bias as the displacement of signal energy in both the CAF and PVT solutions. This process involves not only geometrically projecting multipath errors from individual satellite channels but also considering the combined effects of multiple satellites. Theoretical characterization begins with constructing geometric portraits of multipath errors in CAF and PVT, involving derivations that account for geometric projections relative to satellite elevation and azimuth angles. It was observed that, with a fixed CAF bias, PVT bias increases with higher elevation angles. Therefore, from a geometric perspective, optimal satellite selection strategies for DPE in multipath scenarios entail choosing a configuration that balances high and low elevation angles effectively. Additionally, a satellite circular multipath bias (SCMB) model is introduced, where the receiver's true position serves as the center, the PVT bias of a single satellite as the radius, and the correlation peak centerline as the tangent. Using SCMB, tangent intersection points serve as solutions representing PVT estimates from single satellite channels to multi-satellite superpositions. Lastly, due to the intricate nature of multipath in urban environments, multipath biases in CAF vary across different satellite channels. SCMB is employed to establish upper and lower bounds for PVT errors and to determine the feasible range of PVT solutions in DPE design. The accuracy of the multipath geometric representation is validated through Monte Carlo simulations and urban canyon tests, which can facilitate the future multipath mitigation in urban applications.

The structure of this paper is as follows: Section II introduces the signal models, the geometric multipath models, and the theoretical SCMB model within DPE framework. Section III validates the geometric theory through simulations. Section IV tests the DPE performance through urban experiments with the analysis of signal intensity. Section V concludes the paper and discusses future work.

## II. Geometrical portraits of DPE multipath errors

### A. GNSS signal models

The digital binary phase shift keying (BPSK) modulated GNSS baseband signal, $s[t]$, with multipath can be written as:

$$s[t] = \sum_{m=1}^{M} \sum_{n=0}^{N_m} \left( A_n^m D^m \left[ t - \tau_n^m \right] C^m \left[ t - \tau_n^m \right] \mathrm{e}^{j2\pi f_{dn}^m t + j\varphi_n^m} + \eta_n^m [t] \right) \tag{1}$$

where $t$ denotes the sampling time instant at the sampling rate $f_s$. $m$ represents the satellite index, $m = 1, 2, \cdots, M$ and $M$ is the total number of visible satellites, $n$ represents the signal propagation path index, $n = 0, 1, \cdots, N_m$ and $N_m$ is the total number of propagation path of the $m^{\text{th}}$ GNSS satellite. $n = 0$ denotes the line of sight (LOS) signal. $A_n^m$ represents the signal amplitude, $D^m[\cdot]$ is the navigation message, and $C^m[\cdot]$ is the pseudo-random code. The received code delay, Doppler frequency, and initial phase are modeled by $\tau_n^m$, $f_{dn}^m$, and $\varphi_n^m$. $\eta_n^m[t]$ are the noise term on different paths of multiple different satellite channels, which are identically distributed with the same variance $\sigma_{IF}^2$.

In GNSS receiver, the correlation is conducted for each satellite separately. Therefore, the estimates of signal synchronization parameters, $\hat{\boldsymbol{\nu}}^m = [\hat{\tau}^m, \hat{f}_d^m]$, can be obtained in each channel [14]:

$$\hat{\boldsymbol{\nu}}^m = \arg\max_{\tilde{\tau}^m, \tilde{f}_d^m} \left\{ \sum_{n=0}^{N_m} \mathcal{R}_\tau(\Delta\tau_n^m) \mathcal{R}_f(\Delta f_{dn}^m) \right\} \tag{2}$$

where $\mathcal{R}_\tau(\cdot)$ and $\mathcal{R}_f(\cdot)$ are the correlation function for code delay and Doppler as follows:

$$\mathcal{R}_\tau(\Delta\tau_n^m) \approx \begin{cases} 1 - |\Delta\tau_n^m| & , \ 0 \le |\Delta\tau_n^m| \le 1 \\ 0, & \text{otherwise} \end{cases} \tag{3}$$

$$\mathcal{R}_f\left(\Delta f_{dn}^m\right) = \operatorname{sinc}\left(\pi \Delta f_{dn}^m T_c\right) \tag{4}$$

Here $T_c$ denotes the coherent integration time, and $\Delta\tau_n^m = \tilde{\tau}^m - \tau_n^m$ and $\Delta f_{dn}^m = \tilde{f}_d^m - f_{dn}^m$ are the code delay and Doppler deviations for the $n^{th}$ path, with $(\tilde{\cdot})$ denoting the candidate values of parameters.

Let multipath error defined as $\delta\boldsymbol{\nu}^m = [\delta\tau^m, \delta f_d^m]$. It represents the bias between the signal synchronization parameter estimates, $\hat{\boldsymbol{\nu}}^m = [\hat{\tau}^m, \hat{f}_d^m]$, and the true value of LOS signal, $\boldsymbol{\nu}_0^m = [\tau_0^m, f_{d0}^m]$:

$$\hat{\boldsymbol{\nu}}^m = \boldsymbol{\nu}_0^m + \delta\boldsymbol{\nu}^m \tag{5}$$

where $\delta\tau^m$ and $\delta f_d^m$ are the code delay and Doppler bias for the $m^{\text{th}}$ satellite.

DPE utilizes the following relation between the signal synchronization parameter estimates, $\hat{\boldsymbol{\nu}}^m = [\hat{\tau}^m, \hat{f}_d^m]$ in CAF, and the corresponding PVT parameter estimations, $\hat{\boldsymbol{\gamma}} = [\hat{\mathbf{p}}, \delta\hat{t}, \hat{\mathbf{v}}, \delta\hat{\dot{t}}]$ [6] in navigation domain:

$$\hat{\tau}^m = -\frac{f_c}{c}\left(\|\mathbf{p}^m - \hat{\mathbf{p}}\| + (c\delta\hat{t} - c\delta t^m)\right) \tag{6}$$

$$\hat{f}_d^m = -\frac{f_L}{c}\left(\frac{(\mathbf{p}^m - \hat{\mathbf{p}})^T}{\|\mathbf{p}^m - \hat{\mathbf{p}}\|}(\mathbf{v}^m - \hat{\mathbf{v}}) + \left(c\delta\hat{\dot{t}} - c\delta\dot{t}^m\right)\right) \tag{7}$$

where $c$ is the speed of light, $f_c$ is the PRN code rate, and $f_L$ is the carrier frequency. It is worth noting that this study addresses BPSK-modulated signals without imposing restrictions on the code rate or carrier frequency. By applying different code rates and carrier frequencies for various types of BPSK signals, such as GPS L1, L2C, and L5, the similar analysis can be conducted accordingly. Note that in (6) and (7), $\mathbf{p}^m = [x^m, y^m, z^m]^T$, $\mathbf{v}^m = [\dot{x}^m, \dot{y}^m, \dot{z}^m]^T$, $\delta t^m$ and $\delta\dot{t}^m$ are the position, velocity, clock bias and drift of $m^{\text{th}}$ GNSS satellite. $\hat{\mathbf{p}} = [\hat{x}, \hat{y}, \hat{z}]^T$, $\hat{\mathbf{v}} = \left[\hat{\dot{x}}, \hat{\dot{y}}, \hat{\dot{z}}\right]^T$, $\delta\hat{t}$ and $\delta\hat{\dot{t}}$ are the estimates of receiver. It is worth noting that the position and velocity used in this paper are all defined in East-North-Up (ENU) reference frame.

Theoretically, the MLE of PVT parameters in DPE can be obtained [6]:

$$\hat{\boldsymbol{\gamma}} = \arg\max_{\tilde{\mathbf{p}}, \delta\tilde{t}, \tilde{\mathbf{v}}, \delta\tilde{\dot{t}}}\left\{\left|\sum_{m=1}^{M}\Lambda^m(\tilde{\mathbf{p}}, \delta\tilde{t}, \tilde{\mathbf{v}}, \delta\tilde{\dot{t}})\right|\right\} \tag{8}$$

where $\Lambda^m(\cdot)$ is the CAF expressed by the PVT parameter vector. Similarly with the signal synchronization parameter estimates, the relationship between PVT parameter estimates, $\hat{\boldsymbol{\gamma}}$, and true value, $\boldsymbol{\gamma}$ can be defined as:

$$\hat{\boldsymbol{\gamma}} = \boldsymbol{\gamma} + \delta\boldsymbol{\gamma} \tag{9}$$

where $\delta\boldsymbol{\gamma} = [\delta\mathbf{p}, \delta t_e, \delta\mathbf{v}, \delta\dot{t}_e]$, is the PVT estimate bias including $\delta\mathbf{p} = [\delta x, \delta y, \delta z]$ and $\delta\mathbf{v} = [\delta\dot{x}, \delta\dot{y}, \delta\dot{z}]$.

It has been demonstrated that the code delay and Doppler can be estimated independently with minimal loss of accuracy [15]. Therefore, (2) can be approximately separated into two correlation functions, $\mathcal{R}_\tau(\cdot)$ and $\mathcal{R}_f(\cdot)$, for code delay and Doppler respectively [6], [16], [17].

From (3) and (4), the optimization problems can be solved with low complexity by decoupling the 8-dimensional vector $\tilde{\boldsymbol{\gamma}}$, into two 4-dimensional vectors, $[\tilde{\mathbf{p}}, \delta\tilde{t}]$ and $[\tilde{\mathbf{v}}, \delta\tilde{\dot{t}}]$, which involve:

1) the MLE of position-bias when the velocity-drift are fixed as true value, $\mathbf{v}$ and $\delta\dot{t}$, $\Lambda^m(\tilde{\mathbf{p}}, \delta\tilde{t}, \mathbf{v}, \delta\dot{t}) = \mathcal{R}_\tau(\Delta\tau_n^m)$.

2) the MLE of velocity-drift when the position-bias are fixed as true value, $\mathbf{p}$ and $\delta t$, $\Lambda^m(\mathbf{p}, \delta t, \tilde{\mathbf{v}}, \delta\tilde{\dot{t}}) = \mathcal{R}_f(\Delta f_{dn}^m)$.

## B. Geometrical multipath projection from CAF to PVT

Fig. 1 visualizes the LOS and NLOS signal correlations with the projection from CAF to PVT geometrically. To illustrate this better, one LOS path and one NLOS path are provided for demonstration. Figure 1 (a) and (d) show the CAF of code delay and Doppler, respectively. It is evident that the time delay bias $\delta\tau_n^m$ and Doppler shift bias $\delta f_{dn}^m$ of the NLOS path relative to the LOS path severely deform the composite CAF due to the multipath signals. This deformation is reflected in the PVT solutions as shown in Figure 1 (b) and (c), where the LOS and NLOS correlations overlap and their peak center lines, such as $l_0^m$ and $l_n^m$, are separated by distances $\delta\rho_n^m$ or $\delta\dot{\rho}_n^m$ as seen in Figure 1 (c) and (f). In our previous study, we derived the analytical expression for the LOS center line $l_0^m$, noting that its slope and intercept vary with different satellite elevation and azimuth angles [13]. Based on this theoretical foundation, the NLOS center line $l_n^m$ can be easily obtained, and the distances $\delta\rho_n^m$ and $\delta\dot{\rho}_n^m$ can be derived accordingly. This allows us to trace the multipath errors in PVT solutions all the way back to the CAF bias.

From Fig. 1(a) and (d), the signal parameter estimates bias for the $n^{\text{th}}$ NLOS signal, $\Delta\boldsymbol{\nu}_n^m = [\Delta\tau_n^m, \Delta f_{dn}^m]$, can be expressed by that of the LOS signal, $\Delta\boldsymbol{\nu}_0^m = [\Delta\tau_0^m, \Delta f_{d0}^m]$, as:

$$\Delta\tau_n^m = \Delta\tau_0^m + \delta\tau_n^m \tag{10}$$

$$\Delta f_{dn}^m = \Delta f_{d0}^m + \delta f_{dn}^m \tag{11}$$

By referring to [13], $\Delta\tau_0^m$ and $\Delta f_{d0}^m$ can be expressed using PVT parameters:

$$\Delta\tau_0^m \approx -\frac{f_c}{c}\left(\frac{(\mathbf{p}^m - \mathbf{p})^T}{\|\mathbf{p}^m - \mathbf{p}\|}(\mathbf{p} - \tilde{\mathbf{p}}) + (c\delta t - c\delta\tilde{t})\right) \tag{12}$$

$$\Delta f_{d0}^m = -\frac{f_L}{c}\left(\frac{(\mathbf{p}^m - \mathbf{p})^T}{\|\mathbf{p}^m - \mathbf{p}\|}(\mathbf{v} - \tilde{\mathbf{v}}) + (c\delta\dot{t} - c\delta\tilde{\dot{t}})\right) \tag{13}$$

where $\tilde{\mathbf{p}}$ and $\tilde{\mathbf{v}}$ are the candidate position and velocity, $\delta\tilde{t}$ and $\delta\tilde{\dot{t}}$ are the candidate clock bias and shift. For the sake of clarity, we now consider the $z$ coordinate and receiver clock bias or shift are known or vary slowly with time and can be tracked by other methods [13]. Therefore, (12) and (13) can be transformed as:

$$\Delta\tau_0^m = \frac{f_c}{c \cdot r^m}\left((x^m - x)(x - \tilde{x}) + (y^m - y)(y - \tilde{y})\right) \tag{14}$$

$$\Delta f_{d0}^m = \frac{f_L}{c \cdot r^m}\left((x^m - x)(\dot{x} - \tilde{\dot{x}}) + (y^m - y)(\dot{y} - \tilde{\dot{y}})\right) \tag{15}$$

where $r^m = \sqrt{(x^m - x)^2 + (y^m - y)^2 + (z^m - z)^2}$ is the distance between satellite and receiver.

Subsequently, the CAF results are transformed into PVT parameters and the position and velocity spaces are decoupled using (3) and (4). The 3D view of PVT solutions in Fig. 1(b) and (e) can be projected into 2D view as in Fig. 1(c) and (f) for both position and velocity spaces. It has been proved that the correlation peak center lines, i.e, the $l_0^m$ and $l_n^m$ in Fig. 1 are straight lines [13]. Therefore, the linear geometrical model can be obtained when the code delay and Doppler satisfying:

$$\begin{cases} \Delta\tau_0^m = 0 & l_0^m - \text{LOS} \\ \Delta\tau_n^m = \Delta\tau_0^m + \delta\tau_n^m = 0 \ \ (n \neq 0) & l_n^m - \text{NLOS} \end{cases} \tag{16}$$

$$\begin{cases} \Delta f_{d0}^m = 0 & l_0^m - \text{LOS} \\ \Delta f_{dn}^m = \Delta f_{d0}^m + \delta f_{dn}^m = 0 \ \ (n \neq 0) & l_n^m - \text{NLOS} \end{cases} \tag{17}$$

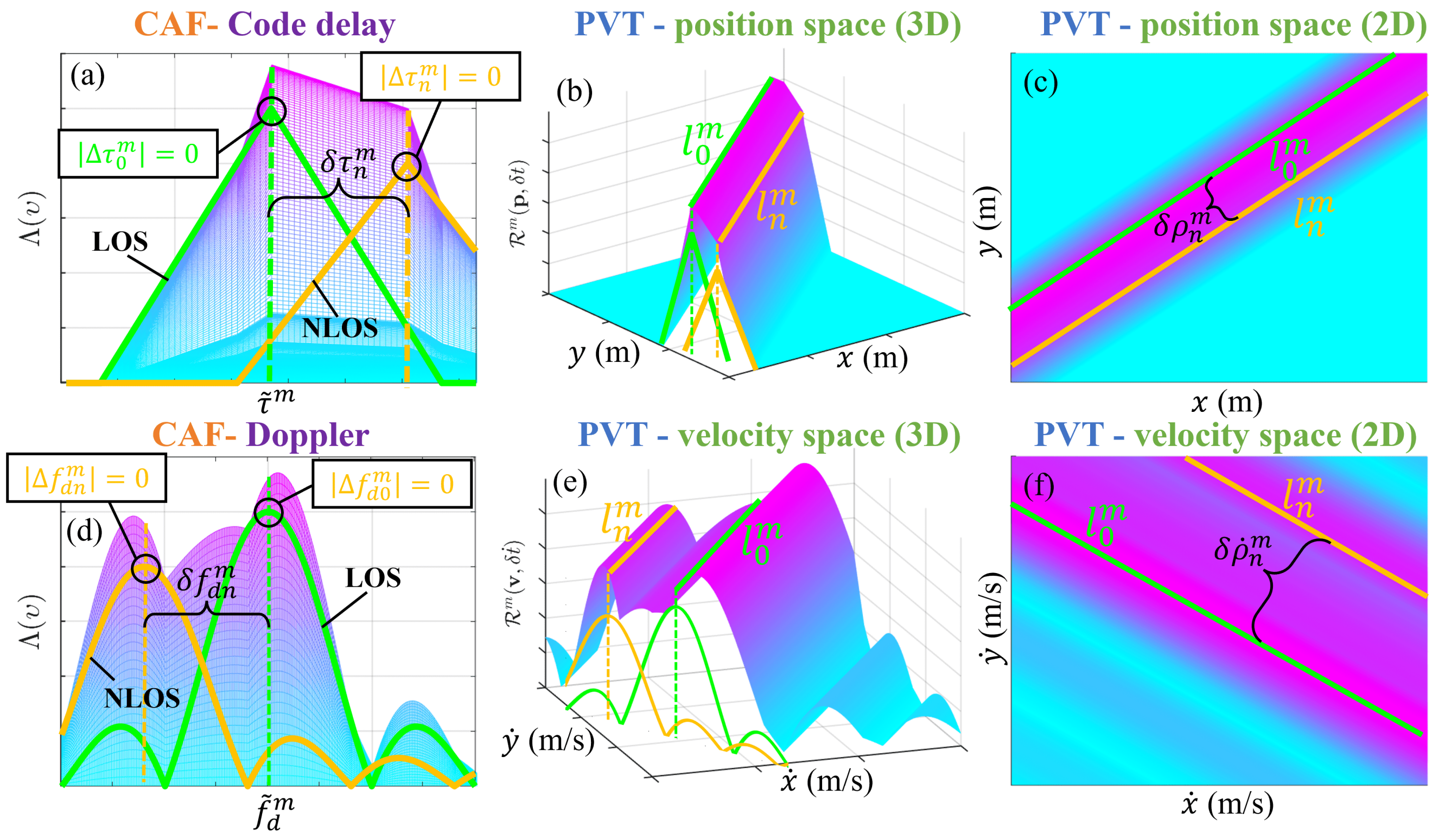

Fig. 1: The geometry characteristic schematic for multipath signals in DPE receiver from the view of (a) CAF in code delay space, (b) position space in 3D, (c) position space in 2D, (d) CAF in Doppler space, (e) velocity space in 3D, and (f) velocity space in 2D.

Then by substituting (14) and (15) into the (16) and (17), the geometric model for the LOS and NLOS can be obtained as linear equations defined by the slope ($k_p^m$ and $k_v^m$) and intercept ($b_{pn}^m$ and $b_{vn}^m$) [13]:

$$\text{position space: } \tilde{y} = k_p^m \tilde{x} + b_{pn}^m \tag{18}$$

$$\text{velocity space: } \tilde{\dot{y}} = k_v^m \tilde{\dot{x}} + b_{vn}^m \tag{19}$$

where the slopes are:

$$\text{position space: } k_p^m = -\frac{x^m - x}{y^m - y} \tag{20}$$

$$\text{velocity space: } k_v^m = -\frac{x^m - x}{y^m - y} \tag{21}$$

and the intercepts are:

$$b_{pn}^m = \begin{cases} y + \dfrac{x^m - x}{y^m - y} x & l_0^m - \text{LOS} \\ y + \dfrac{x^m - x}{y^m - y} x \mp \dfrac{c \cdot r^m \cdot \delta\tau_n^m}{f_c \, (y^m - y)} & l_n^m - \text{NLOS} \end{cases} \tag{22}$$

$$b_{vn}^m = \begin{cases} \dot{y} + \dfrac{x^m - x}{y^m - y} \dot{x} & l_0^m - \text{LOS} \\ \dot{y} + \dfrac{x^m - x}{y^m - y} \dot{x} \mp \dfrac{c \cdot r^m \cdot \delta f_{dn}^m}{f_L \, (y^m - y)} & l_n^m - \text{NLOS} \end{cases} \tag{23}$$

Note that the LOS intercept is attained when $\delta\tau_0^m = 0$ and $\delta f_{d0}^m = 0$. Based on the analysis in [13], the slopes and intercepts are the functions of the elevation $\phi^m$ and azimuth $\theta^m$ as:

$$\text{position space: } \tilde{y} = -\tilde{x} \cdot \tan\theta^m + b_{pn}^m \tag{24}$$

$$\text{velocity space: } \tilde{\dot{y}} = -\tilde{\dot{x}} \cdot \tan\theta^m + b_{vn}^m \tag{25}$$

with the intercepts of [13]:

$$b_{pn}^m = \begin{cases} y + x \cdot \tan\theta^m & l_0^m - \text{LOS} \\ y + x \cdot \tan\theta^m \mp \dfrac{c \cdot \delta\tau_n^m}{f_c \cos\phi^m \cos\theta^m} & l_n^m - \text{NLOS} \end{cases} \quad (26)$$

$$b_{vn}^m = \begin{cases} \dot{y} + \dot{x} \cdot \tan\theta^m & l_0^m - \text{LOS} \\ \dot{y} + \dot{x} \cdot \tan\theta^m \mp \dfrac{c \cdot \delta f_{dn}^m}{f_L \cos\phi^m \cos\theta^m} & l_n^m - \text{NLOS} \end{cases} \quad (27)$$

As shown in Fig. 1(c) and (f), there are multipath-induced range biases between the correlation peak of LOS signal and NLOS signals, denoted by the $\delta\rho_n^m$ and $\delta\dot{\rho}_n^m$ in the position and velocity space. Based on the theory of algebraic geometry, the multipath bias can be calculated using the formula for the distance between parallel lines. In position space, the range bias, $\delta\rho_n^m$, can be calculated using (24) and (26) as:

$$\delta\rho_n^m = \frac{|b_{pn} - b_{p0}|}{\sqrt{\tan^2\theta^m + 1}} = \frac{c}{f_c}\delta\tau_n^m \sec\phi^m \quad (28)$$

It shows that the multipath bias, $\delta\rho_n^m$, in position space is the geometrical projection of code delay bias $\delta\tau_n^m$ in CAF through elevation $\phi^m$. It is independent of satellite azimuth $\theta^m$. The changes in azimuth $\theta^m$ result in variations in the slope $k_p^m$, see (24).

Similarly, the multipath bias $\delta\dot{\rho}_n^m$ in velocity space can be obtained through (25) and (27):

$$\delta\dot{\rho}_n^m = \frac{|b_{vn} - b_{v0}|}{\sqrt{\tan^2\theta^m + 1}} = \frac{c}{f_L}\delta f_{dn}^m \sec\phi^m \quad (29)$$

It means $\delta\dot{\rho}_n^m$ in velocity space is the geometrical projection of Doppler bias $\delta f_{dn}^m$ in CAF through elevation $\phi^m$ but independent of azimuth $\theta^m$.

In essence, (28) and (29) emphasize the propagation of multipath bias from CAF to PVT when focusing on the correlation peak in horizontal projection. The larger the code delay or Doppler bias in CAF, the lager bias in PVT. Meanwhile, the higher elevation angle results in lager PVT bias. Consequently, satellites at lower elevations present more favorable horizontal geometric characteristics in PVT. However, as analyzed in our previous work [13], lower-elevation satellites may introduce higher noise variance. Therefore, satellite selection in DPE should consider a balanced assessment of factors including signal strength and geometric characteristics.

## C. Satellite circular multipath bias model

Based on the analysis above, it can be observed that there are similarities between position space and velocity space. Therefore, in the subsequent analysis, we will take position space as an example for derivation. The results can similarly be applied to velocity space.

Fig. 2 illustrates the range bias $\delta\rho_n^m$ relative to different satellite azimuths $\theta^m$. When considering the receiver position truth **p** as the central reference, the range bias $\delta\rho_n^m$ depends solely on the multipath code delay $\delta\tau_n^m$ and the elevation angle $\phi^m$, as expressed in (28). Importantly, $\delta\rho_n^m$ remains unaffected by azimuth $\theta^m$, implying its constancy across different azimuths. This constancy results in the perpendicular geometric projection forming a circular trajectory around the truth **p**, with a radius of $\delta\rho_n^m$. It is important to note that while azimuth $\theta^m$ does not affect $\delta\rho_n^m$, it determines the slope $k_p^m$ along the arc tangent direction as in Fig. 2(a). As a result, to visualize the range bias $\delta\rho_n^m$ with the geometrical satellite distribution, the satellite circular multipath bias (SCMB) model in Fig. 2(b) is proposed.

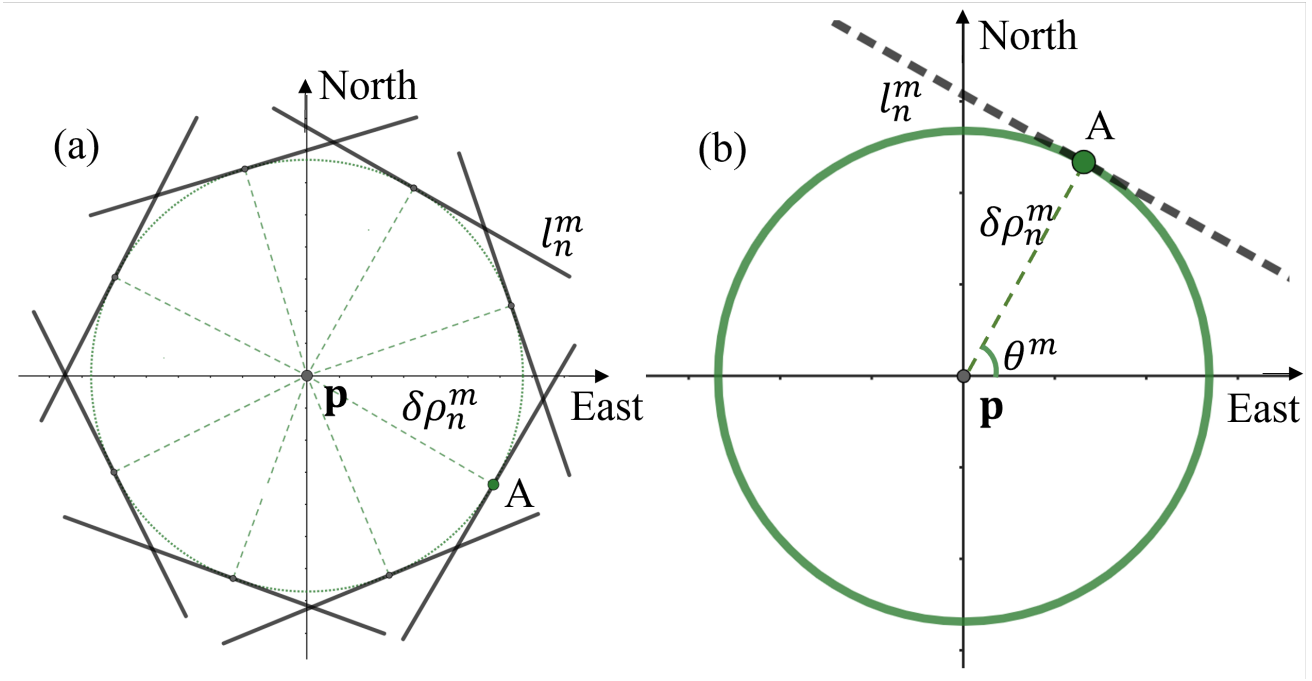

Fig. 2: Illustration of the range bias $\delta\rho_n^m$ relative to the diverse satellite azimuths $\theta^m$ using (a) circular trajectory of the perpendicular geometric projection. (b) single satellite circular multipath bias model

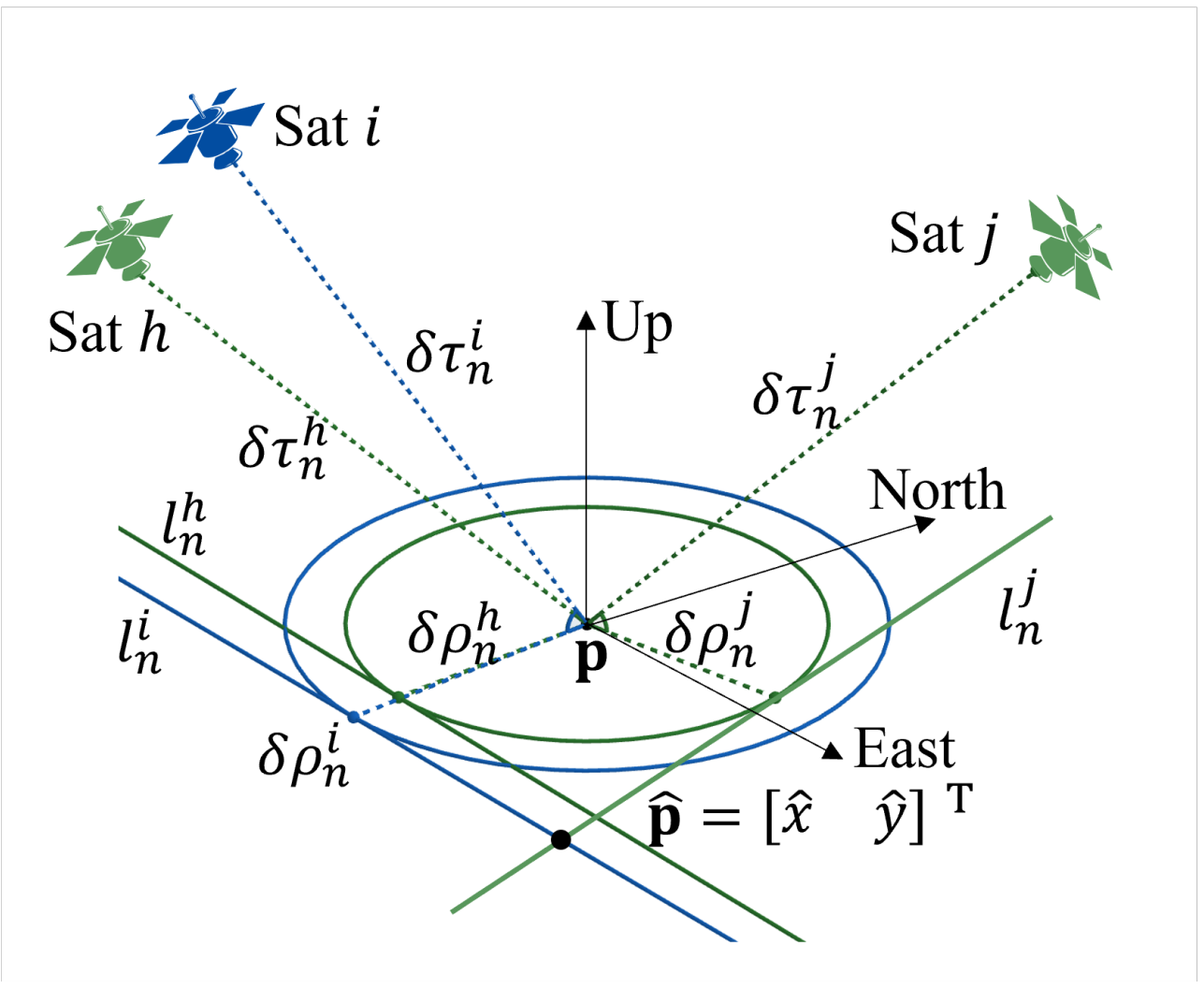

Fig. 3: Illustration of the range bias $\delta\rho_n^m$ relative to various elevations $\phi^m$ using the SCMB.

Fig. 3 illustrates the range bias $\delta\rho_n^m$ relative to various elevations $\phi^m$ using the SCMB. In Fig. 3, three satellites exhibit distinct geometric distributions: Satellites $h$ and $i$ share the same azimuth, while Satellites $h$ and $j$ share the same elevation, with Sat $i$ having a higher elevation angle than the others. Assuming these satellites experience the same multipath delay, the projected range bias $\delta\rho_n^i$ of Satellite $i$ will be larger than that of Satellites $h$ and $j$, due to the inverse relationship with elevation $\phi^m$ as described in (28). According to SCMB, Satellites $h$ and $j$ define identical inner bias circles with equal radii ($\delta\rho_n^h = \delta\rho_n^j$), while Satellite $i$ defines an outer circle with a larger radius $\delta\rho_n^i$. Furthermore, because Satellites $h$ and $i$ share the same azimuth, according to (24), the slopes, $k_p^h$ and $k_p^i$, of the correlation peak center lines, $l_n^h$ and $l_n^i$, are identical, indicating they are parallel. In contrast, Satellites $h$ and $j$ share the same elevation angle but differ in azimuth; thus, their respective circular trajectories have the same radius, but their correlation peak center lines are not parallel.

In the DPE framework, the estimated PVT solution $\hat{\mathbf{p}}$ is determined by superposing the correlation peaks obtained from signals transmitted by multiple satellites. Visualizing this in a 2D plane, the intersection of correlation peaks from signals of two different satellites provides a position estimate, denoted as $\hat{\mathbf{p}}$, see Fig. 3. However, in real-world scenarios affected by multipath interference, each satellite may contribute multiple intersection points on the plane. These points represent potential positions where the receiver could be located. DPE selects the PVT solution $\hat{\mathbf{p}}$ that corresponds to the intersection point with the highest correlation strength.

In the presence of various multipath delays, the estimated PVT solution $\hat{\mathbf{p}}$ can deviate from the true position due to multipath effects. To accurately characterize these errors, it is crucial to account for the multipath induced range bias $\delta\rho_n^m$ associated with each satellite. This bias reflects the discrepancy between the actual range measurement and the measured range affected by multipath interference. In error propagation analysis within the DPE framework, the multipath induced range bias $\delta\rho_n^m$ for each satellite is converted to PVT deviation through the following analysis.

Given the expressions of the correlation peak lines $l_n^i$ and $l_n^j$ of satellites $i$ and $j$ from (24) and (26):

$$\begin{cases} \tilde{y} = -\tilde{x} \cdot \tan\theta^i + b_{pn}^i & \text{Satellite } i \\ \tilde{y} = -\tilde{x} \cdot \tan\theta^j + b_{pn}^j & \text{Satellite } j \end{cases} \tag{30}$$

The intersection point between $l_n^i$ and $l_n^j$ yields a position estimate $\hat{\mathbf{p}} = [\hat{x}, \hat{y}]$ as in Fig. 3. This point satisfies both equations in (30), therefore we have:

$$\begin{cases} \hat{x} = \dfrac{b_{pn}^i - b_{pn}^j}{\tan\theta^i - \tan\theta^j} \\ \\ \hat{y} = \dfrac{b_{pn}^i \tan\theta^j - b_{pn}^j \tan\theta^i}{\tan\theta^i - \tan\theta^j} \end{cases} \tag{31}$$

Substituting the intercept (26) into (31), the multipath induced position estimation error can be obtained:

$$\begin{cases} \delta x = \hat{x} - x = \dfrac{c}{f_c} \dfrac{\delta\tau_n^j \sec\phi^j \cos\theta^i - \delta\tau_n^i \sec\phi^i \cos\theta^j}{\sin(\theta^j - \theta^i)} \\ \\ \delta y = \hat{y} - y = \dfrac{c}{f_c} \dfrac{\delta\tau_n^j \sec\phi^j \sin\theta^i - \delta\tau_n^i \sec\phi^i \sin\theta^j}{\sin(\theta^j - \theta^i)} \end{cases} \tag{32}$$

In (32), a mapping relationship from the multipath delay, $\delta\tau_n^i$ and $\delta\tau_n^j$, to the position estimate bias, $\delta x$ and $\delta y$, is provided. The mapping function links the geometric parameters, e.g., azimuth and elevation, between the CAF bias and PVT bias.

By substituting (28) into (32), the direct relationship from the range bias on each satellite, e.g., $\delta\rho_n^i$ and $\delta\rho_n^j$, to the positioning bias ($\delta x$ and $\delta y$) can be constructed as well:

$$\begin{cases} \delta x = \dfrac{\delta\rho_n^j \cos\theta^i - \delta\rho_n^i \cos\theta^j}{\sin(\theta^j - \theta^i)} \\ \\ \delta y = \dfrac{\delta\rho_n^j \sin\theta^i - \delta\rho_n^i \sin\theta^j}{\sin(\theta^j - \theta^i)} \end{cases} \tag{33}$$

Let $\Delta\theta = \theta^j - \theta^i$ ($\theta^j \neq \theta^i$) denotes the azimuth difference between satellites $i$ and $j$ with 0 in minimum and $2\pi$ in maximum. Then the horizontal position estimation error $\delta r$ can be expressed as:

$$\delta r = \sqrt{(\delta x)^2 + (\delta y)^2} = \frac{\sqrt{(\delta\rho_n^i)^2 + (\delta\rho_n^j)^2 - 2\delta\rho_n^i\delta\rho_n^j \cos\Delta\theta}}{\sin\Delta\theta} \tag{34}$$

Note that this distance error $\delta r$ exhibits geometric symmetry when $\Delta\theta$ is between $[0, \pi]$ and $[\pi, 2\pi]$. To keep $\delta r$ positive, we restrict the range of $\Delta\theta$ to $[0, \pi]$. From the perspective of geometric matching, this limited range represents the vector angle between the LOS observations from the two satellites at the receiver.

Similarly, based on (29), the multipath estimation error in velocity space can be expressed as $\delta\dot{x}$, $\delta\dot{y}$ and $\delta\dot{r}$ will be:

$$\begin{cases} \delta\dot{x} = \dfrac{\delta\dot{\rho}_n^j \cos\theta^i - \delta\dot{\rho}_n^i \cos\theta^j}{\sin(\theta^j - \theta^i)} \\ \\ \delta\dot{y} = \dfrac{\delta\dot{\rho}_n^j \sin\theta^i - \delta\dot{\rho}_n^i \sin\theta^j}{\sin(\theta^j - \theta^i)} \\ \\ \delta\dot{r} = \dfrac{\sqrt{(\delta\dot{\rho}_n^i)^2 + (\delta\dot{\rho}_n^j)^2 - 2\delta\dot{\rho}_n^i\delta\dot{\rho}_n^j \cos\Delta\theta}}{\sin\Delta\theta} \end{cases} \tag{35}$$

The algebraic derivation presented above illustrates that the estimation errors in PVT solution, denoted as $\delta r$ and $\delta\dot{r}$ in (34) and (35), are influenced by the range and range rate biases from each satellite, as well as by the azimuthal distribution of the satellites. Notably, when examining equations (34) and (35), the denominators adhere to the cosine law. Linking this observation with our proposed SCMB model, we can visualize geometrically how these PVT estimation errors manifest.

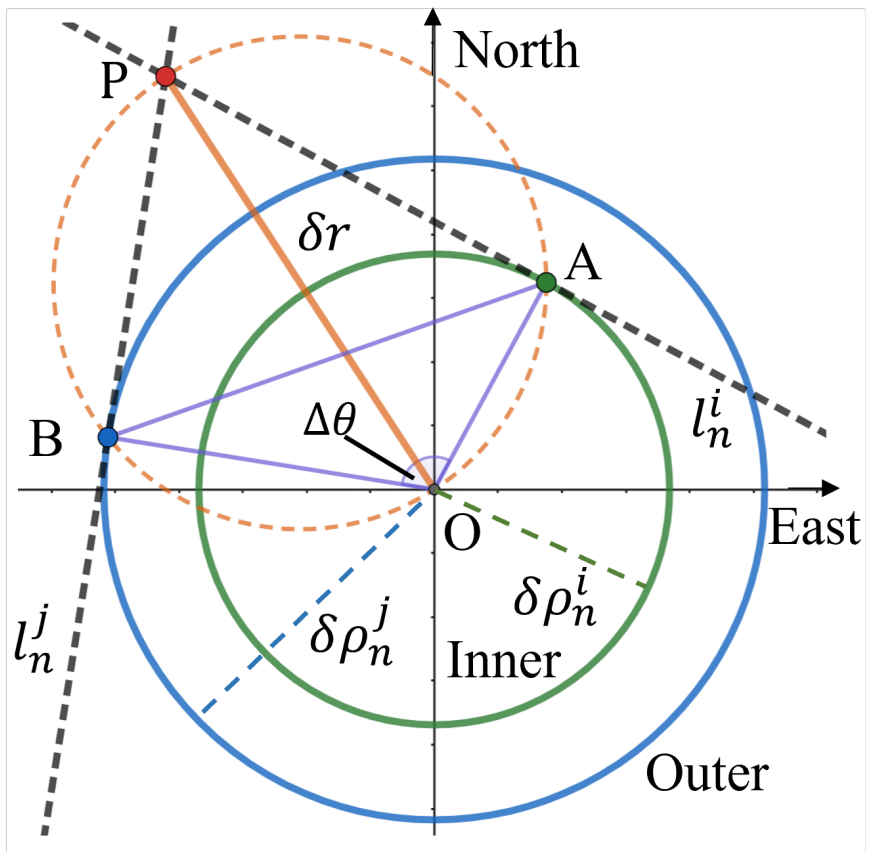

Fig. 4: Geometrical portrait of multipath error in position solution $\delta r$ through the proposed SCMB model.

Fig. 4 illustrates a schematic diagram of the SCMB model featuring two satellites. As previously mentioned, the circle radius corresponds to the range biases $\delta\rho_n^i$ and $\delta\rho_n^j$. Points A and B represent the tangency points where the correlation peak center lines intersect with the satellite circles. Point P denotes the estimated position of the receiver $\hat{\mathbf{p}}$, determined by the intersection of the two correlation peak center lines $l_n^i$ and $l_n^j$. Within the triangle $\triangle$AOB, the angle $\angle$AOB denotes the azimuthal difference $\Delta\theta$ between two satellites. Apparently, using the cosine rule, the length of AB side, |AB|, can be expressed as:

$$|\mathrm{AB}| = \sqrt{(\delta\rho_n^i)^2 + (\delta\rho_n^j)^2 - 2\delta\rho_n^i\delta\rho_n^j \cos\Delta\theta} \quad (36)$$

Comparing (36) to (34), it can be observed that (34) can be expressed geometrically in terms of the length of |AB|:

$$\delta r = \frac{|\mathrm{AB}|}{\sin\Delta\theta} \quad (37)$$

Based on (37), we can encompass $\triangle$AOB within a circumscribed circle, where the diameter **OP** represents the positioning error $\delta r$, as derived from the sine rule. Based on the cyclic quadrilateral theorem, for a cyclic quadrilateral, the sum of its opposite angles is $\pi$, which means the angle between two lines ($l_n^i$ and $l_n^j$), $\angle\mathrm{APB} = \pi - \Delta\theta$. Since the range of $\Delta\theta$ is $[0, \pi]$, the range of $\angle$APB is also $[0, \pi]$. Thus, examining changes in the length of $OP$ offers a geometric perspective on the positioning error. Such a geometric portrait enhances the intuitive understanding and facilitates the analysis of PVT multipath errors.

For the sake of generality, we define the angle between two lines $l_n^i$ and $l_n^j$ as $\Delta\alpha$. According to (24) and (25), $\Delta\alpha$ can be represented by $\Delta\theta$ as follows:

$$\Delta\alpha = \tan^{-1}\left|\frac{\tan\theta^j - \tan\theta^i}{1 + \tan\theta^j\tan\theta^i}\right| = \begin{cases}\Delta\theta, & \Delta\theta \in [0, \frac{\pi}{2}] \\ \pi - \Delta\theta, & \Delta\theta \in [\frac{\pi}{2}, \pi]\end{cases} \quad (38)$$

From the geometric definition it is known that $\Delta\alpha \in [0, \pi/2]$, therefore, $\Delta\alpha$ is a supplementary angle of $\Delta\theta$ when $\Delta\theta$ is obtuse.

## D. Multipath error bound in PVT solution

The expression $\delta r$ in (37) is influenced by both the range bias $\delta\rho_n^m$ and the azimuthal difference $\Delta\theta$. The varied geometric distribution of azimuths results in different outcomes for $\delta r$. To establish bounds on multipath errors in PVT estimations, whether minimum $\delta r_{min}$ or maximum $\delta r_{max}$, it is essential to calculate the partial derivative of $\delta r$ in (34) as follows from an algebraic perspective:

$$\begin{cases}\dfrac{\partial\delta r}{\partial\Delta\theta} = \dfrac{\delta\rho_n^i\delta\rho_n^j\cos^2\Delta\theta - ((\delta\rho_n^i)^2 + (\delta\rho_n^j)^2)\cos\Delta\theta + 1}{\left(\sqrt{(\delta\rho_n^i)^2 + (\delta\rho_n^j)^2 - 2\delta\rho_n^i\delta\rho_n^j\cos\Delta\theta}\right)\sin^2\Delta\theta} \\ \dfrac{\partial\delta r}{\partial\delta\rho_n^i} = \dfrac{\delta\rho_n^i - \delta\rho_n^j\cos\Delta\theta}{\left(\sqrt{(\delta\rho_n^i)^2 + (\delta\rho_n^j)^2 - 2\delta\rho_n^i\delta\rho_n^j\cos\Delta\theta}\right)\sin\Delta\theta} \\ \dfrac{\partial\delta r}{\partial\delta\rho_n^j} = \dfrac{\delta\rho_n^j - \delta r^i\cos\Delta\theta}{\left(\sqrt{(\delta\rho_n^i)^2 + (\delta\rho_n^j)^2 - 2\delta\rho_n^i\delta\rho_n^j\cos\Delta\theta}\right)\sin\Delta\theta}\end{cases} \quad (39)$$

Thus, setting the partial derivative (39) to zero allows us to identify the critical points of the multipath position estimate bias. Without loss of generality, we assume $\delta\rho_n^j > \delta\rho_n^i$ in the following context:

$$\frac{\partial\delta r}{\partial\delta\theta} = 0,\ \frac{\partial\delta r}{\partial\delta\rho_n^i} = 0,\ \frac{\partial\delta r}{\partial\delta\rho_n^j} = 0 \Rightarrow \begin{cases}\Delta\theta = \cos^{-1}\dfrac{\delta\rho_n^i}{\delta\rho_n^j} \\ \delta\rho_n^i = \delta\rho_n^j\cos\Delta\theta\end{cases} \quad (40)$$

By substituting (40) into (34), the extreme values of $\delta r$, i.e., $\delta r_{min}$ or $\delta r_{max}$, can be determined. (40) shows a symmetrical effect of $\delta\rho_n^i$ and $\delta\rho_n^j$. It is interesting to note that $\delta r$ is a concave function of $\Delta\theta$ within $[0, \pi]$. When $\Delta\theta \to 0$ or $\pi$, $\delta r \to \infty$. As $\Delta\theta$ increases from 0 to $\cos^{-1}(\delta\rho_n^i/\delta\rho_n^j)$, $\delta r$ gradually decreases from $\infty$ to its minimum at $\delta\rho_n^j$. It means that the minimum positioning error $\delta r_{min}$ depends on the satellite with the larger multipath range error. When $\Delta\theta$ continues to increase, e.g., from $\cos^{-1}(\delta\rho_n^i/\delta\rho_n^j)$ to $\pi$, $\delta r$ increases from $\delta\rho_n^j$ to its maximum at $\infty$ again.

The algebraic analysis results can be explained geometrically. Fig. 5 illustrates the three critical nodes of $\delta r$ as $\Delta\theta$ changes from 0 to $\pi$. In this figure, the center O represents the receiver location, while points

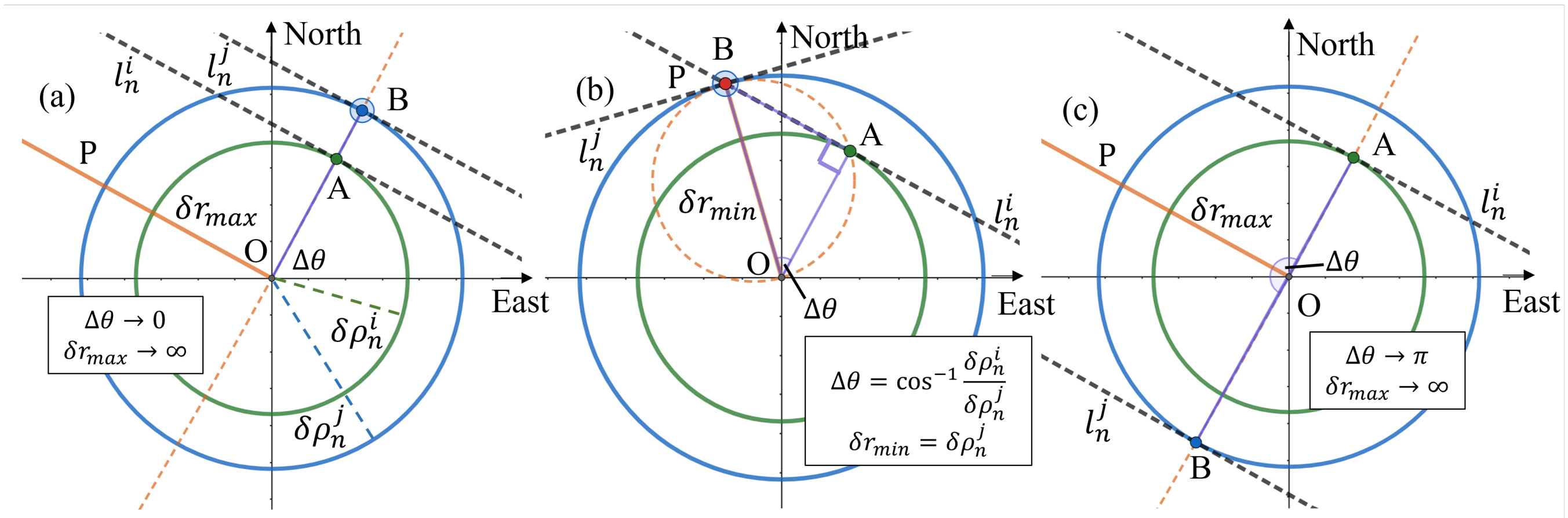

Fig. 5: Multipath error bound in position estimation with respect to (a) $\Delta\theta \to 0$, (b) $\Delta\theta = \cos^{-1}(\delta\rho_n^i/\delta\rho_n^j)$ and (c) $\Delta\theta \to \pi$ by using the SCMB model.

A and B denote the azimuthal projections of the two visible satellites in the E-N plane. For instance, when $\Delta\theta = 0$, the two satellites lie in the same direction, causing OA and OB in Fig. 5(a) to overlap. In this scenario, the signal projection lines $l_n^i$ and $l_n^j$ are nearly parallel, indicating that the intersection point between the two lines is infinitely distant, resulting in $\delta r_{\max} = \infty$. As satellite $j$'s azimuth moves from the OA direction to the OB direction, the value of $\Delta\theta$ increases. In Fig. 5(b), when $\Delta\theta$ increases to $\cos^{-1}(\delta\rho_n^i/\delta\rho_n^j)$, the triangle AOB becomes a right triangle. Here, the intersection point between OA and OB represents the position estimate with a minimum error of $\delta\rho_n^j$. As $\Delta\theta$ continues to increase to $\pi$, as shown in Fig. 5(c), the two satellites are symmetrically distributed on both sides of center O. In this case, the signal projection lines from the two satellites are nearly parallel again, causing the position estimate error $\delta r$ to tend towards infinity once more.

In summary, when $\delta\rho_n^i$ and $\delta\rho_n^j$ are fixed, e.g., $\delta\rho_n^j > \delta\rho_n^i$, the minimum value of the position estimate error $\delta r$ equals to the larger multipath range error $\delta\rho_n^j$, while the maximum value tends to infinity. This analysis provides a reference for analyzing the impact of individual satellite multipath bias on the final DPE positioning result. The similar conclusion can be made for velocity estimation. To extend the above analysis to a more general case involving multiple satellites with multiple signal paths, we will examine all potential combinations of multipath errors and theoretically establish PVT error bounds.

Assuming there is total $M$ visible satellites and each with $N_m$ different signal paths, the range error vector can be denoted as $\boldsymbol{\delta\rho} = [\delta\rho_0^1, \delta\rho_1^1, \cdots, \delta\rho_{N_1}^1, \cdots, \delta\rho_n^m, \cdots, \delta\rho_{N_M}^M]$.

For this case, using the SCMB model, there will have multiple intersection points, see Fig. 6. The intersection of the pairwise produces the intersection points, according to permutation combination, there will be $0.5\left(\sum_{m=1}^M N_m\right)^2 - 0.5\sum_{m=1}^M N_m^2$ intersection points in total.

For convenient illustration, each satellite only present one path either in LOS or NLOS. The final PVT estimate error can be simply categorized into three special scenarios based on diverse multipath environment [18]: 1) most satellites are in LOS path with only one satellite in NLOS path, e.g., $\boldsymbol{\delta\rho} = [0, ..., \delta\rho_n^m, ..., 0]$, 2) all satellites with equal NLOS range bias $\delta\rho$, e.g., $\boldsymbol{\delta\rho} = [\delta\rho, ..., \delta\rho, ..., \delta\rho]$, and 3) all satellites with different NLOS range errors, e.g., $\boldsymbol{\delta\rho} = [\delta\rho_1^1, \cdots, \delta\rho_n^m, \cdots, \delta\rho_{N_M}^M]$ $(n \neq 0)$. As shown in Fig. 6, we take four satellites, numbered with #1, #2, #3, and #4, as an example to introduce these three cases, where the angles between the lines, $\Delta\alpha$, are marked accordingly.

1) Case 1 - only one satellite has NLOS range bias: When only Satellite #1 has a NLOS signal and the remaining satellites all have LOS signals, there are visually four intersection points, $P_1$, $P_2$, $P_3$, and $P_4$, representing the possible DPE solutions, as shown in Fig. 6(a). All the LOS signals intersect at point $P_1$, which corresponds to the receiver's true location at the circle center O. The only NLOS signal from Satellite #1 forms a circle with the radius of $\delta\rho_n^m$. The rest points represent possible position solutions with different estimation errors, or equivalently the different error distances $|OP_2|$, $|OP_3|$, and $|OP_4|$. In this case, the only non-zero range error $\delta\rho$ originates from that NLOS path. It can be observed in Fig. 6(a) that the smaller $\Delta\alpha$ corresponds to the longer error distances, i.e., the larger estimation error. Therefore, based on the relationship between $\Delta\theta$ and $\Delta\alpha$ in (38), the position estimation error can be written as follow from (34):

$$\delta r = \frac{\delta\rho_n^m}{\sin\Delta\theta} \tag{41}$$

And $\delta r$ can reach to infinity in maximum theoretically. While the minimum position error is taken at a specific

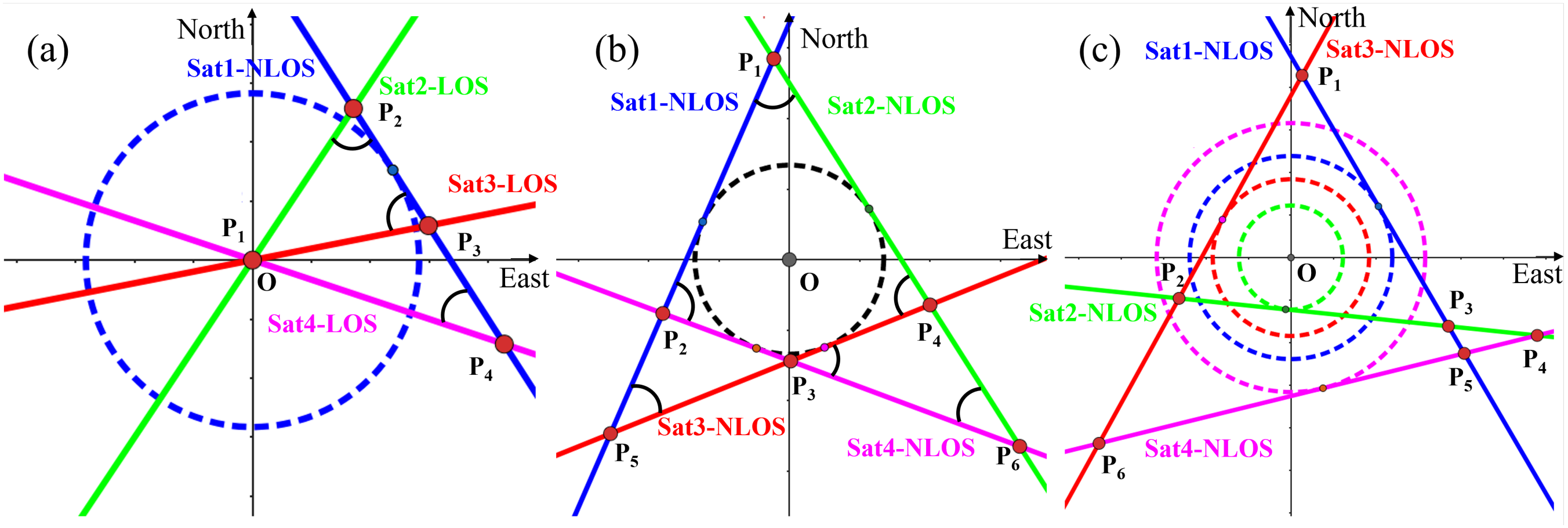

Fig. 6: Schematic of three special scenarios when (a) only one satellite has a nonzero NLOS bias, (b) similar NLOS bias for all satellites and (c) different NLOS bias for different satellites

value of $\Delta\theta = \pi/2$, which is $\delta r_{min} = \delta\rho_n^m$. The multipath error bound for DPE position estimation is $\delta r \in [\delta\rho_n^m, \infty)$.

Similarly, for the velocity space, the velocity error $\delta\dot{r}$ can be simplified as:

$$\delta\dot{r} = \frac{\delta\dot{\rho}_n^m}{\sin\Delta\theta} \tag{42}$$

The multipath error bound for velocity estimation error is $\delta\dot{r} \in [\delta\dot{\rho}_n^m, \infty)$.

2) Case 2 - equal NLOS range bias $\delta\rho$ for all satellites: When all satellites have the same non-zero range error, $\delta\rho$, see Fig. 6(b), the geometrical multipath portrait for all satellites shares the same concentric circle with the same radius of $\delta\rho$. All satellite correlation peak lines are tangents to the multipath circle, which means $\delta\rho_n^i = \delta\rho_n^j = \delta\rho$ in (34). In this case, it can be observed in Fig. 6(b) that the smaller $\Delta\alpha$ also corresponds to the longer error distances, i.e., the larger estimation error. As responded by Fig. 6(b), it can be observed that the smaller $\Delta\alpha$, the larger $\delta r$, and $\delta r$ is close to infinity when $\Delta\alpha$ is near 0 theoretically. While the minimum position error is achieved as the two lines approach coincidence.

Based on the relationship between $\Delta\theta$ and $\Delta\alpha$ in (38) and substituting $\delta\rho$ into (34), then we have:

$$\delta r = \frac{\sqrt{2\delta\rho^2(1-\cos\Delta\theta)}}{\sin\Delta\theta} = \frac{2\sqrt{\delta\rho^2\sin^2\frac{\Delta\theta}{2}}}{2\sin\frac{\Delta\theta}{2}\cos\frac{\Delta\theta}{2}} = \delta\rho\sec\frac{\Delta\theta}{2} \tag{43}$$

That is when $\Delta\theta \to 0$, $\delta r_{min} \to \delta\rho$. The multipath error bound for DPE position estimation is $\delta r \in (\delta\rho, \infty)$.

Similarly, for the velocity space, the velocity error $\delta\dot{r}$ can be simplified as:

$$\delta\dot{r} = \delta\dot{\rho}\sec\frac{\Delta\theta}{2} \tag{44}$$

The multipath error bound for velocity estimation error is $\delta\dot{r} \in (\delta\dot{\rho}, \infty)$.

3) Case 3 - different $\delta\rho_n^m$ for each satellite: When all satellites have different non-zero NLOS errors, i.e., $\delta\boldsymbol{\rho} = [\delta\rho_1^1, ..., \delta\rho_n^m, ..., \delta\rho_{N_M}^M]$ $(n \neq 0)$, the position estimate error $\delta r$ needs to comprehensively consider the deviation of the correlation domain of each satellite and the azimuth difference $(\Delta\theta)$ between each pair of satellites. The position estimate bias range can be judged by combining (34) and Fig. 5.

Combining Fig.5 and Fig. 6(c), it can be observed that the $\delta r$ can also reach to infinity in maximum theoretically. While the minimum position error $\delta r$ depends on the second smallest range bias. When $\Delta\theta = \cos^{-1}(\delta\rho_n^i/\delta\rho_n^j)$, $\delta r_{min} = \delta\rho_n^j$, where $\delta\rho_n^i$ and $\delta\rho_n^j$ are the smallest and the second smallest values in $[\delta\rho_1^1, ..., \delta\rho_n^m, ..., \delta\rho_{N_M}^M]$ respectively. The multipath error bound for DPE position estimation is $\delta r \in [\delta\rho_n^j, \infty)$.

Similarly, for the velocity space, the multipath error bound is $\delta\dot{r} \in [\delta\dot{\rho}_n^j, \infty)$, where $\delta\dot{\rho}_n^j$ is the second smallest values in $[\delta\dot{\rho}_1^1, ..., \delta\dot{\rho}_n^m, ..., \delta\dot{\rho}_{N_M}^M]$.

## E. Theoretical analysis on geometrical multipath error propagation

From the theoretical derivation above, it can be observed that the elevation angle $\phi^m$ primarily affects the error mapping from CAF biases $(\delta\tau_n^m, \delta f_{dn}^m)$ to range measurement errors $(\delta\rho_n^m, \delta\dot{\rho}_n^m)$. In contrast, the azimuth angle $\theta^m$ mainly influences the error mapping from range measurements $(\delta\rho_n^m, \delta\dot{\rho}_n^m)$ to PVT solutions $(\delta r, \delta\dot{r})$. Here, we summarize the impact of the elevation angle $\phi^m$ and the azimuth angle $\theta^m$ on error propagation.

Given a NLOS path delay of $\delta\tau_n^m = 1$ chip and a Doppler shift of $\delta f_{dn}^m = 120$ Hz on the GPS L5 signal, where $f_c = 10.23$ MHz and $f_L = 1176.45$ MHz, the measurement errors on range $\delta\rho_n^m$ and range rate $\delta\dot{\rho}_n^m$ for different elevation angles $\phi$ can be determined according to (28) and (29).

Fig. 7 clearly illustrates the inversely relationship between multipath errors and satellite geometric elevation distribution. As the elevation angle $\phi$ increases, the range and range rate errors also increase. The resulting $\delta\rho_n^m$ and $\delta\dot{\rho}_n^m$ can grow from nearly 29.3 m and 30.6 m/s towards $\infty$ as $\phi$ varies from 0°to 90°. Notably, when $\phi$ exceeds approximately 30°, corresponding to a typical elevation cut-off mask, the error growth becomes particularly rapid.

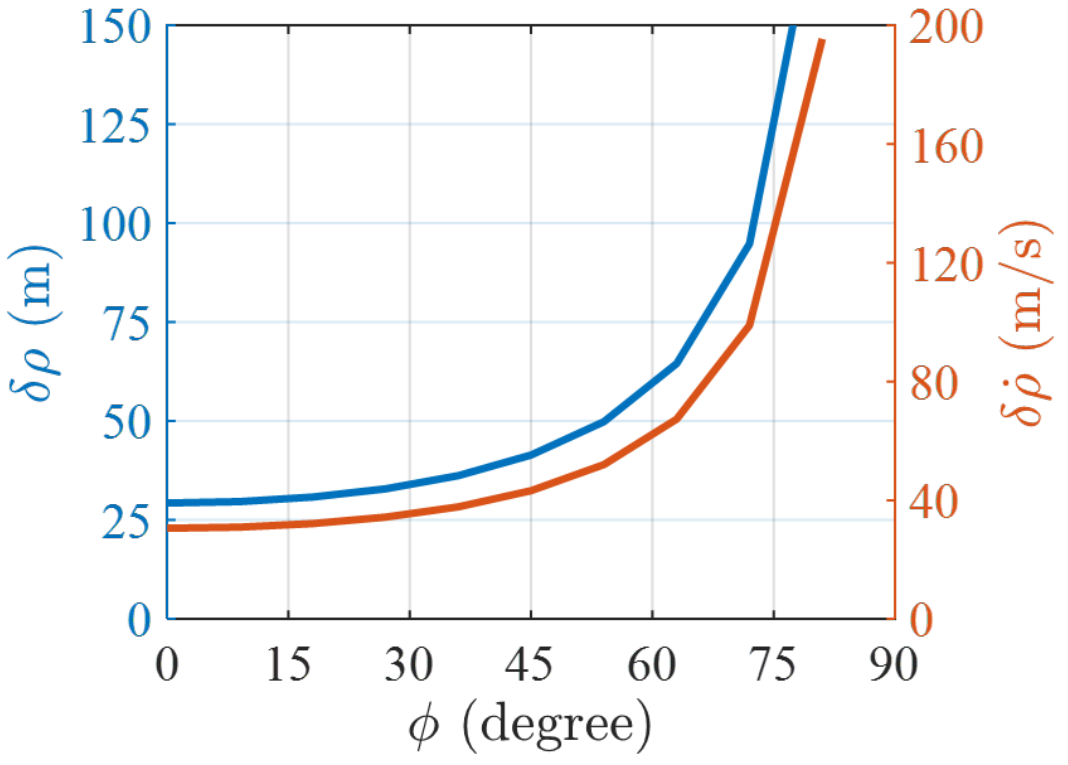

Fig. 7: The variation of $\delta\rho_n^m$ and $\delta\dot{\rho}_n^m$ with respect to the elevation angle $\phi$ when there is a NLOS delay of 1 chip and Doppler shift of 120 Hz on GPS L5 signal.

Assuming that only one satellite has specific measurement errors in range and range rate ($\delta\rho$, $\delta\dot{\rho}$) as discussed in Case 1, or that measurement errors in range and range rate ($\delta\rho$, $\delta\dot{\rho}$) are equally present in all visible satellites as introduced in Case 2, the impact of the azimuth angle $\theta^m$ is depicted in Fig. 8.

In Case 2, the values of PVT solution errors $\delta r$ and $\delta\dot{r}$ increase monotonically with increasing $\Delta\theta$, whereas in Case 1, they form a concave function of $\Delta\theta$. The minimum values of $\delta r$ and $\delta\dot{r}$ are $\delta\rho$ and $\delta\dot{\rho}$ when $\Delta\theta$ = 0° or 90° in Cases 1 and 2, respectively.

## III. Simulation Validation

In this section, the DPE multipath error characterization using geometrical propagation will be validated through Monte Carlo simulation, where the multipath error propagation process from multipath delay $\delta\tau_n^m$ and Doppler shift $\delta f_{dn}^m$ to range and range rate measurement errors ($\delta\rho_n^m$, $\delta\dot{\rho}_n^m$) in single satellite channel, then to the PVT estimation errors of $\delta r$ and $\delta\dot{r}$ in DPE with multiple satellite composition, will be validated and analyzed. The geometrical impacts of the satellite elevation angle $\phi^m$ and azimuth angle difference $\Delta\theta$ will be examined accordingly.

### A. Simulation setup

Table I outlines the simulation parameters, including the true position and velocity coordinates of the receiver in the ECEF coordinate system, based on a driving test conducted on September 3, 2022, in the Shanghai Lujiazui urban environment. For illustrative purposes, this receiver coordinate is used as the origin point O and converted to the ENU coordinate system. The simulation incorporates four satellites: PRN#10, #18, #23, and #24, with their positions determined from the ephemeris data. The elevation and azimuth angles of these satellites are detailed in Table II.

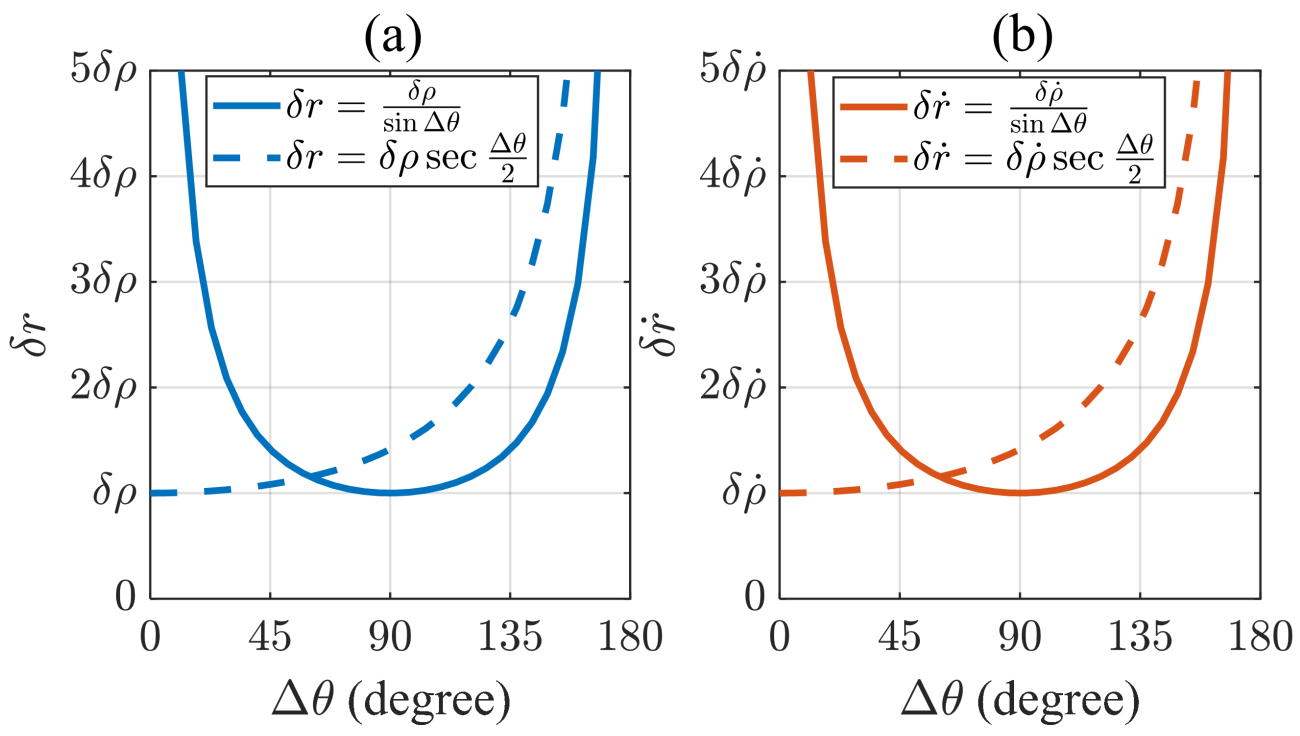

Fig. 8: The PVT solution errors variations with respect to azimuth difference $\Delta\theta$ for (a) $\delta r$ and (b) $\delta\dot{r}$.

TABLE I: Simulation Parameters

| Parameter | Value |
|---|---|
| Data and Time | 2022/9/3 UTC 8:4:47 |
| Receiver Position $\mathbf{p}$ (m) | [-2851838,4653607,3289209] |
| Receiver Velocity $\mathbf{v}$ (m/s) | [-5.5, -4.7, 1.9] |
| Satellite PRN | 10, 18, 23, 24 |
| Code Frequency $f_c$ | 10.23 MHz |
| Carrier Frequency $f_L$ | 1176.45 MHz |
| Sampling Frequency $f_s$ | 30.69 MHz |
| Coherent Integration Time $T_c$ | 20 ms |
| Search Range for $\mathbf{p}$ and $\mathbf{v}$ | ±100 m, ±100 m/s |
| Search Grid Step for $\mathbf{p}$ and $\mathbf{v}$ | 1 m, 0.1 m/s |
| Multipath Delay $\delta\tau_n^m$ | 1 chip |
| Multipath Doppler Shift $\delta f_{dn}^m$ | 120 Hz |
| Range/Range Rate Bias $\delta\rho_n^m/\delta\dot{\rho}_n^m$ for PRN#10,18,23,24 | 60m, 40m, 30m, 15m<br>60m/s, 40m/s, 30m/s, 15m/s |

In the simulation, the GPS L5 signal with a carrier frequency of $f_L$ = 1176.45 MHz and a code frequency of $f_c$ = 10.23 MHz is utilized. As detailed in Table I, correlation results are generated within a PVT search grid, with a position range of ±100 m and a velocity range of ±100 m/s, using step sizes of 0.25 m and 0.25 m/s, respectively. The coherent integration time, $T_c$, is set to 20 ms, aligning with the DPE platform implementation in [4], though the proposed theory is also valid for high-dynamic signal processing with shorter integration times, such as 1 ms. For consistency with the experiment, the sampling rate is set to 30.69 MHz, as specified by the LabSat intermediate frequency collector.

TABLE II: Satellite geometrical distribution

| | Elevation $\phi^m$ (°) | Azimuth $\theta^m$ (°) |
|---|---|---|
| PRN#10 | 35.4 | 320.2 |
| PRN#18 | 42.8 | 213.8 |
| PRN#23 | 66.7 | 336.1 |
| PRN#24 | 69.8 | 45.1 |

The error mapping process in our simulation is divided into two steps. First, we map multipath delay and Doppler shift ($\delta\tau_n^m$, $\delta f_{dn}^m$) to the range and range rate biases ($\delta\rho_n^m$, $\delta\dot{\rho}_n^m$). Second, we map the range and range rate biases to the position and velocity estimate errors ($\delta r$, $\delta\dot{r}$). Simulation parameters are based on actual field test data. For the first mapping step, we set $\delta\tau_n^m$ = 1 chip and $\delta f_{dn}^m$ = 120 Hz for four satellites to assess the impact of elevation angle $\phi^m$. In the second step, we set range/range rate biases for satellites PRN #10, #18, #23, and #24 to 60, 40, 30, 15 m and 60, 40, 30, 15 m/s, respectively. Monte Carlo simulations and analysis in three cases, shown in Fig. 6, evaluate how the azimuth difference angle $\Delta\theta$ affects the theoretical PVT error bounds.

### B. Multipath error propagation from ($\delta\tau_n^m$, $\delta f_{dn}^m$) to ($\delta\rho_n^m$, $\delta\dot{\rho}_n^m$) through geometrical elevation projection

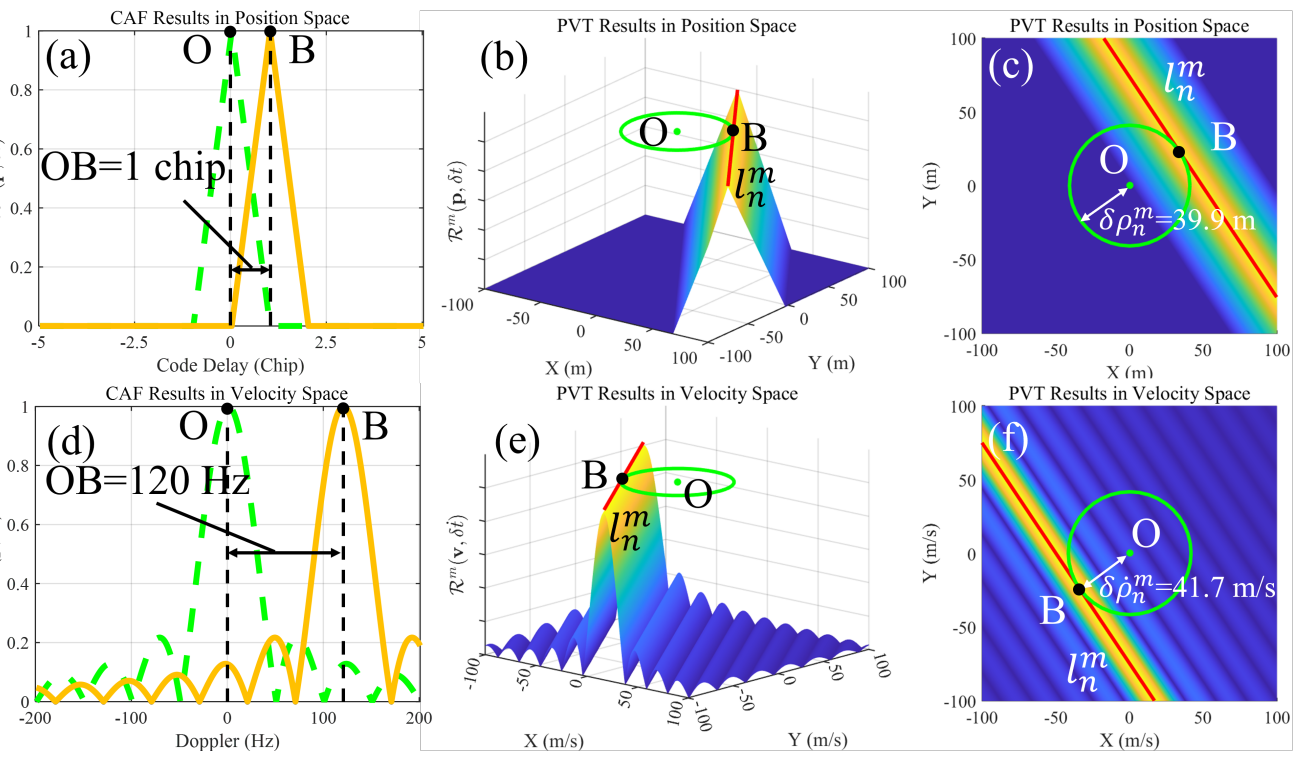

Fig. 9: The simulated NLOS signals CAF with $\delta\tau_n^m$ = 1chip (a) and $\delta f_{dn}^m$ = 120Hz (d) on PRN#18 for the theoretical error propagation on the range (b)-(c) and range rate (e)-(f) errors in 3D and 2D view.

Using PRN#18 as an example, Fig. 9 displays the CAF generated for a multipath delay of 1 chip and a Doppler shift of 120 Hz for a pure NLOS signal. Given the true position O, the geometrical mapping from CAF to range measurements is performed with an elevation angle of 42.8°. In the position space, with a multipath delay of 1 chip in the CAF shown in Fig. 9(a), the range bias |OB| measured in Fig. 9(c) and the theoretical $\delta\rho_n^m$ calculated using (28) are both 39.9 m. Similarly, for a Doppler shift of 120 Hz in Fig. 9(d), the range rate bias |OB| in Fig. 9(f) and the theoretical $\delta\dot{\rho}_n^m$ from (29) are both 41.7 m/s. This agreement verifies the theoretical accuracy of the elevation projection as given in (28) and (29).

TABLE III: The theoretical and simulated elevation projection on range and range rate biases from $\delta\tau_n^m$ = 1 chip and $\delta f_{dn}^m$ = 120 Hz

| | $\delta\rho_n^m$ (m) Theoretical/Simulated | $\delta\dot{\rho}_n^m$ (m/s) Theoretical/Simulated |
|---|---|---|
| PRN#10 | 36.0/36.0 | 37.5/37.5 |
| PRN#18 | 39.9/39.9 | 41.7/41.7 |
| PRN#23 | 74.0/74.0 | 77.2/77.2 |
| PRN#24 | 84.8/84.8 | 88.5/88.5 |

Table III presents both simulated and theoretical results for the four satellites, demonstrating consistency across different elevation and azimuth geometrical distributions, which further validates the theoretical models. Notably, as shown in Table II, the elevation angles of satellites PRN#10 through PRN#24 progressively increase. Correspondingly, for the same multipath delay $\delta\tau_n^m$ and Doppler shift $\delta f_{dn}^m$, the range/range rate biases ($\delta\rho_n^m$, $\delta\dot{\rho}_n^m$) also increase, aligning with the trend in Fig. 7. This confirms the theoretical conclusion that a larger code delay or Doppler shift in CAF results in greater range biases and a more degraded PVT solution.

### C. Multipath error propagation from ($\delta\rho_n^m$, $\delta\dot{\rho}_n^m$) to ($\delta r$,$\delta\dot{r}$) through geometrical azimuth projection

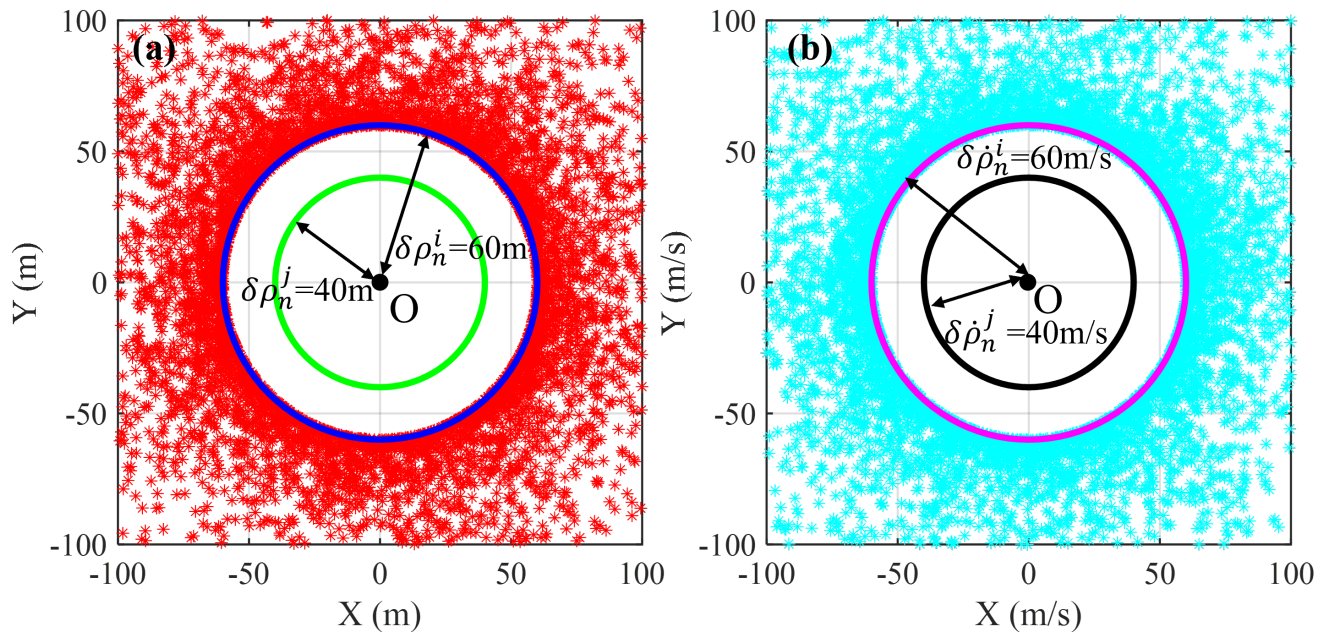

Fig. 10: Monte Carlo verification on PVT errors ($\delta r$,$\delta\dot{r}$) with ($\delta\rho_n^m$, $\delta\dot{\rho}_n^m$) as (60m, 60m/s) and (40m, 40m/s) from PRN#10 and PRN#18 in (a) position space and (b) velocity space.

This subsection further verifies the multipath error propagation from range and range rate biases ($\delta\rho_n^m$, $\delta\dot{\rho}_n^m$) to position and velocity errors ($\delta r$, $\delta\dot{r}$) using geometrical azimuth projection. Assuming the four satellites exhibit different range and range rate biases as listed in Table I, the resulting PVT errors $\delta r$ and $\delta\dot{r}$ are determined using the proposed SCMB model. In this model, the DPE position estimation in a two-dimensional plane is derived from the intersection of correlation peaks from two satellites. When the range biases $\delta\rho_n^i$ and $\delta\rho_n^j$ are fixed, with $\delta\rho_n^j > \delta\rho_n^i$, the minimum position error $\delta r$ is equal to the larger multipath range error $\delta\rho_n^j$, while the maximum value tends toward infinity. This conclusion is validated through Monte Carlo simulations.

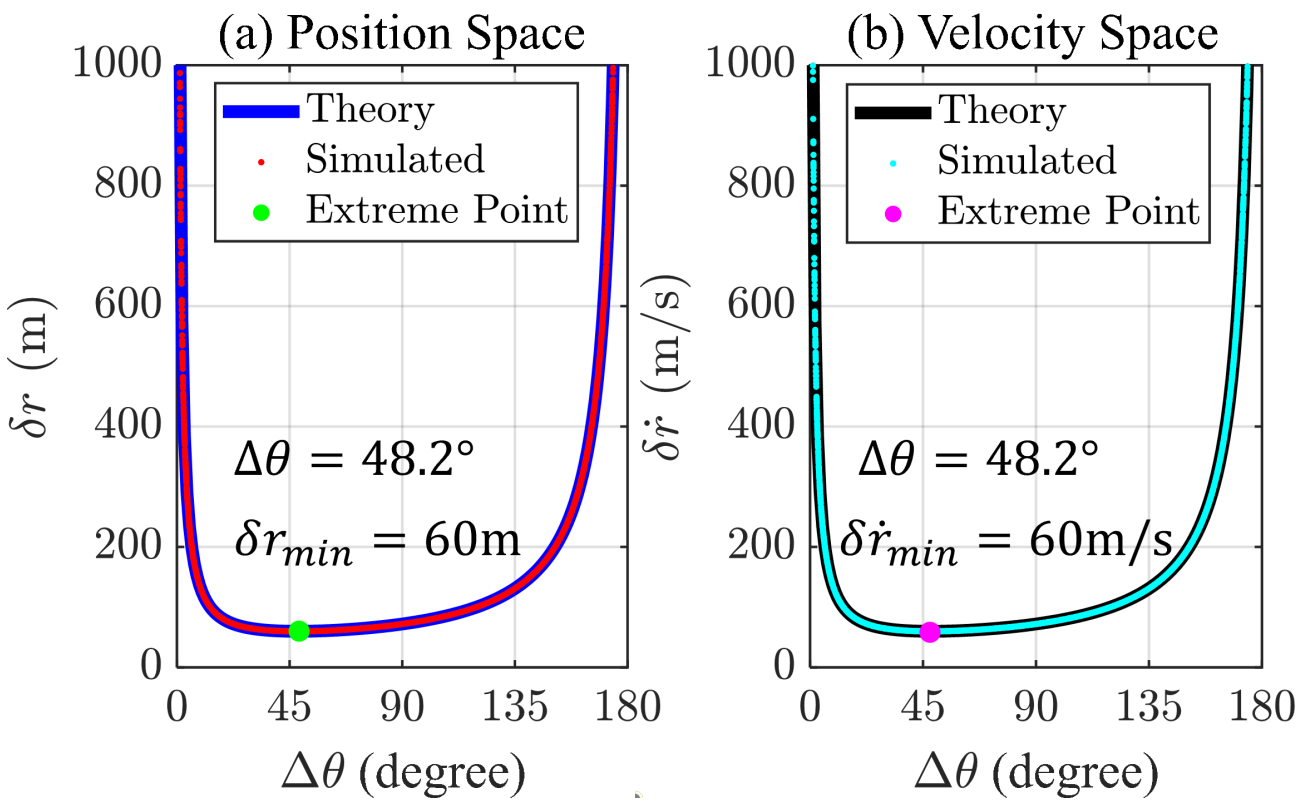

Fig. 11: Monte Carlo verification of $(\delta r, \delta\dot{r})$ as a function of $\Delta\theta$ with given $(\delta\rho_n^m,\ \delta\dot{\rho}_n^m)$ from PRN#10 and PRN#18 as (60m, 60m/s) and (40m, 40m/s), respectively.

For illustration, consider PRN#10 and PRN#18. Their azimuth difference, $\Delta\theta$, is 106.4° as shown in Table II. With the given range and range rate biases, $\delta\rho_n^m$ and $\delta\dot{\rho}_n^m$ from Table I, the resulting PVT errors $\delta r$ and $\delta\dot{r}$ should be 84.4 m. To further validate this theory, we performed a Monte Carlo simulation by randomly generating $\Delta\theta$ uniformly distributed within the range $[0, \pi]$ and repeated the simulation 10,000 times. The PVT errors obtained from this simulation are illustrated in Fig. 10. According to the proposed SCMB, PRN#10 and PRN#18 form two concentric circles around the receiver's true position $O$ in Fig. 10, each representing different radii corresponding to $\delta\rho_n^m$ and $\delta\dot{\rho}_n^m$, respectively. As $\Delta\theta$ varies, the PVT errors $\delta r$ and $\delta\dot{r}$ continuously change. The distribution of the PVT solutions ranges from the radius of the outer circle to infinity, with the distribution shown in Fig. 10 scaled to ±100m and ±100m/s. More specifically, the distance between the PVT solutions and the center of the circle O represents the PVT errors $\delta r$ and $\delta\dot{r}$ in Fig. 10(a) and (b). It is evident that the PVT estimates are situated outside the outer satellite circle, indicating that the PVT errors are always at least as large as the radius of the satellite circle with the larger range or range rate bias $\delta\rho_n^m$, and $\delta\rho_n^m$.

For theoretical verification, the PVT errors $\delta r$ and $\delta\dot{r}$ as functions of $\Delta\theta$ from the Monte Carlo simulation are illustrated in Fig. 11. Additionally, the theoretical values of $\delta r$ and $\delta\dot{r}$ derived from (34) and (35) are provided as a baseline for comparison.

TABLE IV: Theoretical and simulated position estimate biases $\delta r$ in three cases (Unit: m)

| | \|OA\| | \|OB\| | \|OC\| | \|OD\| | \|OE\| |
|---|---|---|---|---|---|
| | Theo./Sim. | Theo./Sim. | Theo./Sim. | Theo./Sim. | Theo./Sim. |
| PRN | 10, 24 | 18, 23 | 10, 18 | 23, 24 | 10, 23 |
| $\Delta\theta$ (°) | 85.1* | 122.3 | 106.4 | 69.0* | 15.9 |
| Case 1 | / | 47.3/46.7 | 41.7/41.9 | / | / |
| Case 2 | 54.2/54.3 | 82.9/82.3 | 66.7/66.8 | 48.5/48.2 | 40.4/40.7 |
| Case 3 | 60.8/61.2 | 72.8/73.2 | 84.4/84.9 | 30.3/30 | / |

* denotes the $\Delta\theta$ has been mapped to the range of $[0, \pi]$.

TABLE V: Theoretical and simulated velocity estimate biases $\delta\dot{r}$ in three cases (Unit: m/s)

| | \|OA\| | \|OB\| | \|OC\| | \|OD\| | \|OE\| |
|---|---|---|---|---|---|
| | Theo./Sim. | Theo./Sim. | Theo./Sim. | Theo./Sim. | Theo./Sim. |
| PRN | 10, 24 | 18, 23 | 10, 18 | 23, 24 | 10, 23 |
| $\Delta\theta$ (°) | 85.1* | 122.3 | 106.4 | 69.0* | 15.9 |
| Case 1 | / | 47.3/46.7 | 41.7/41.9 | / | / |
| Case 2 | 54.2/54.3 | 82.9/82.3 | 66.7/66.8 | 48.5/48.2 | 40.4/40.7 |
| Case 3 | 60.8/61.2 | 72.8/73.2 | 84.4/84.9 | 30.3/30 | / |

* denotes the $\Delta\theta$ has been mapped to the range of $[0, \pi]$.

It show that both $\delta r$ and $\delta\dot{r}$ exhibit a concave relationship with $\Delta\theta$. These errors reach their minimum at $\Delta\theta = \cos^{-1}(40/60) = 48.2°$, corresponding to the relatively larger range and range rate bias from PRN#10 at 60 m and 60 m/s, respectively. This finding aligns with the theoretical predictions presented in (34), (35), and Fig. 5, thereby confirming the validity of the error propagation from $(\delta\rho_n^m, \delta\dot{\rho}_n^m)$ to $(\delta r, \delta\dot{r})$ through geometric projection of $\Delta\theta$.

### D. Multipath error bound analysis in PVT solutions

This part analyzes the multipath error bound in PVT solutions from superimposed projection using multiple satellites. In response to the theoretical analysis in Section II-D, three scenarios in Fig. 6 are simulated, where 1) Case 1: Only PRN#18 is an NLOS signal, introducing a range and range rate bias of 40 m and 40 m/s, respectively. 2) Case 2: All four satellites are assumed to have the same range and range rate biases, both set to 40 m and 40 m/s. 3). Case 3: The four satellites exhibit different range and range rate biases as specified in Table I. Note that the elevation and azimuth of these four satellites are the same as in the Table II.

The CAF peaks and their projections in position and velocity spaces are illustrated in Figs. 12 and 13, respec-tively. In these figures, the circle center O represents the receiver's true position. The circle's radius is determined by the range or range rate biases from the NLOS signals. The various projection lines correspond to the signal geometrical projections at different elevation and azimuth angles, with intersection points such as A, B, C, D, and E representing the PVT candidate estimates.

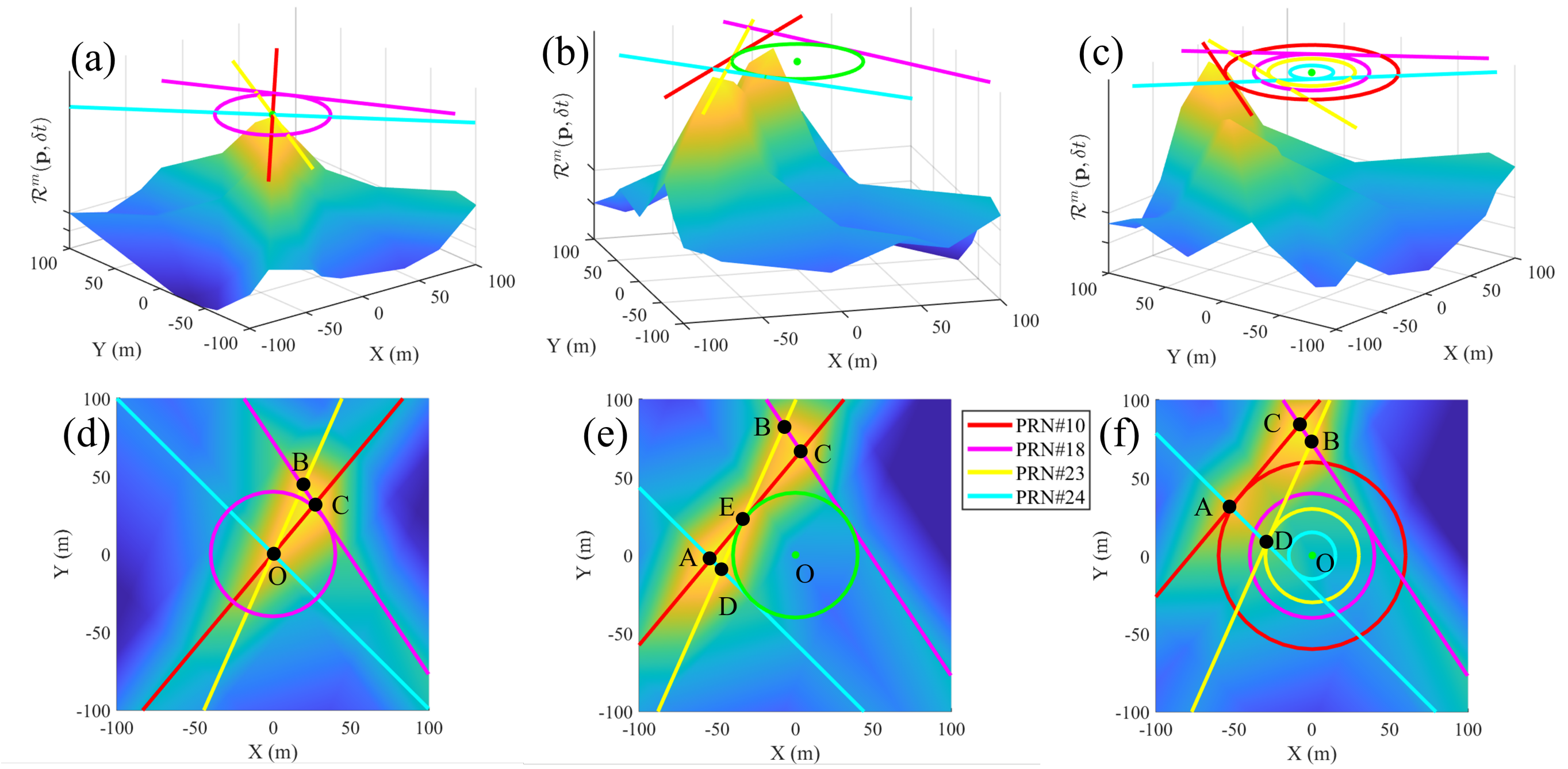

Fig. 12: The multi-satellite correlation values and geometrical projections in 2D and 3D position spaces for theoretical validations in Case 1 (a)(d), Case 2 (b)(e), and Case 3 (c)(f).

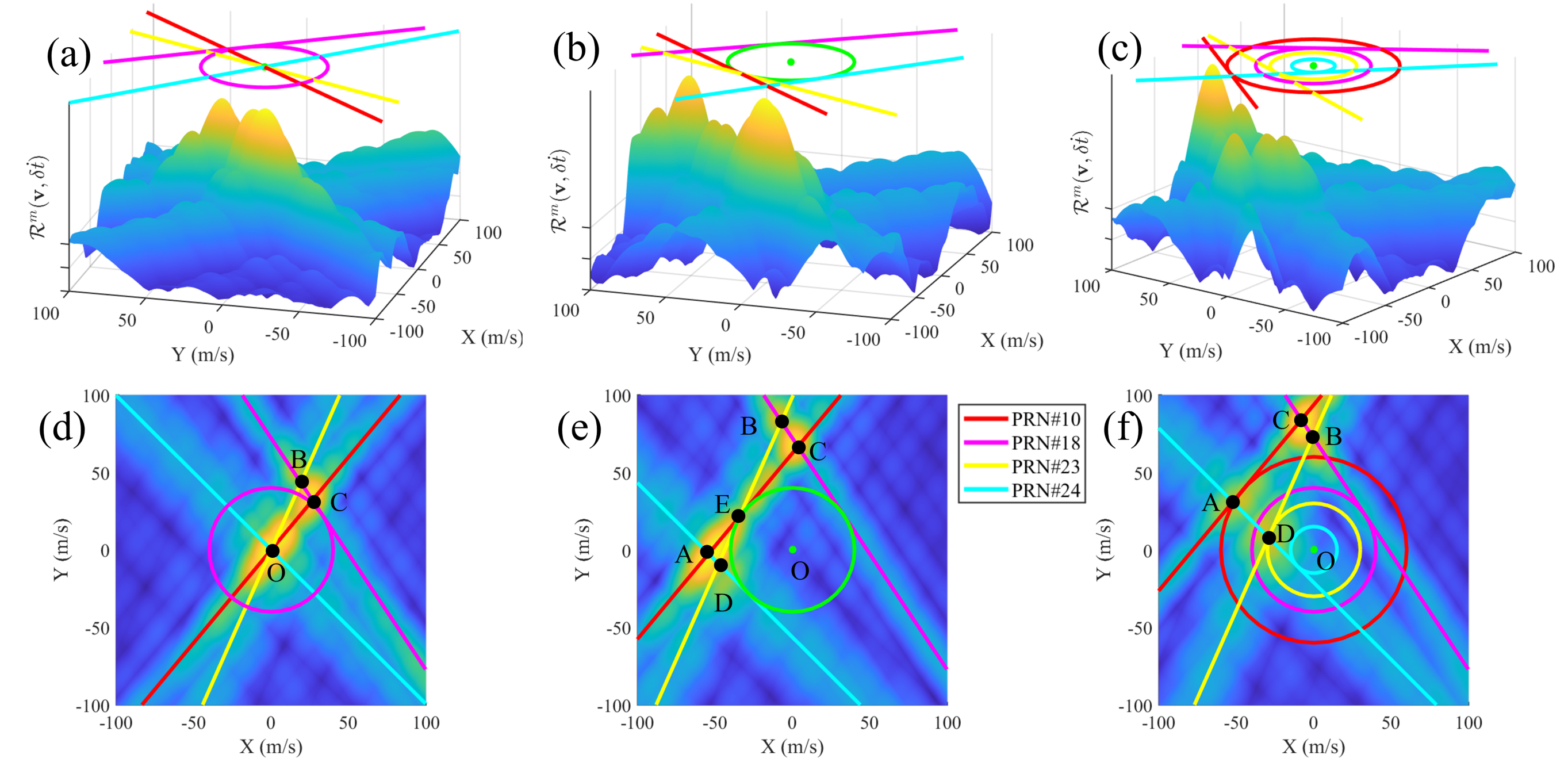

Fig. 13: The multi-satellite correlation values and geometrical projections in 2D and 3D velocity spaces for theoretical validations in Case 1 (a)(d), Case 2 (b)(e), and Case 3 (c)(f).

In Case 1, where only PRN#18 is an NLOS signal, the projected signal lines from PRNs#10, #23, and #24 intersect at the true receiver position O. The NLOS projected lines from PRN#18 are tangent to the circle at radii of 40 m and 40 m/s, as shown in Figs. 12(a)(d) and 13(a)(d). In Case 2, where all four satellites have the same range and range rate biases, their projected lines intersect at the same circle but at different directions due to varied azimuth distributions, as depicted in Figs. 12(b)(e) and 13(b)(e). In Case 3, with satellites exhibiting different range and range rate biases, their projected lines are tangent to different concentric circles with varying radii, as illustrated in Figs. 12(c)(f) and 13(c)(f).

Since the intersection points A, B, C, D, and E in Figs. 12 and 13 representing the candidate PVT estimates, the multipath errors in PVT solutions, $\delta r$ and $\delta \dot{r}$, can be directly obtained by measuring the distance between the receiver truth O to those intersection points. Comparably, the theoretical $\delta r$ and $\delta \dot{r}$ can be calculated from (34) and (35) by substituting the values of $\delta \rho_n^m$ and $\delta \dot{\rho}_n^m$, as well as the azimuth difference $\Delta\theta$ from Table II. Both the theoretical and simulated PVT errors along with $\Delta\theta$ in three cases are listed in Table IV and Table V, respectively. The results demonstrate a strong consistency between the theoretical predictions and the simulation data, further confirming the validity of the geometric projection theory.

For the PVT error bound analysis across different cases: 1).in Case 1, The points B and C represent the intersections of PRN#18/PRN#23 and PRN#10/PRN#18, respectively. According to Table IV, the $\Delta\theta$ for PRN#10/PRN#18 is 106.4°, which is closer to $\pi/2$ compared to the $\Delta\theta$ for PRN#18/PRN#23. Additionally, OC is nearer to the range bias $\delta \rho_n^m$ of 40m for PRN#18. This observation confirms the error bound for Case 1 as described in (41) and Fig. 8, which states that as $\Delta\theta$ approaches $\pi/2$, the error $\delta r$ decreases, with the minimum error $\delta r$ equal to the range bias $\delta \rho_n^m$ of the NLOS signal. 2). In Case 2, as $\Delta\theta$ decreases, the error $\delta r$ also decreases. At point E, where $\Delta\theta$ for PRN#10/PRN#23 is the smallest among the five points, the error $\delta r$ represented by OE is close to the range bias $\delta\rho$ of 40m for Case 2. This verifies the error bound for Case 2 as given in (43) and Fig. 8, which states that the error $\delta r$ increases monotonically with $\Delta\theta$, and when $\Delta\theta = 0$, the minimum error $\delta r$ equals the range bias $\delta\rho$. 3). In Case 3, the positioning error $\delta r$ at point OD is the smallest and closely matches the range bias $\delta \rho_n^m$ of 30 meters for PRN#23. This observation supports the conclusion that "the minimum position error depends on the second smallest range bias."

## IV. Field test

As mentioned in Section III, field test conducted on September 3, 2022, in the Lujiazui area of Shanghai—a typical urban canyon environment—were used for performance validation. GPS raw data of 200s was collected using a high-gain Trimble Zephyr 3 Base Antenna and a LabSat 3 Wideband receiver, with a 0 MHz intermediate frequency (IF), 2-bit quantization, and a sampling rate of 30.69 MHz [19]. The data recording start time and visible satellites during the drive test were consistent with those listed in Table I in Section III. This dataset was analyzed in [10] using a multi-frequency DPE receiver design. For illustration, this work focuses on the GPS L5 signal due to its pronounced multipath features. Additional comparisons can be found in [10].

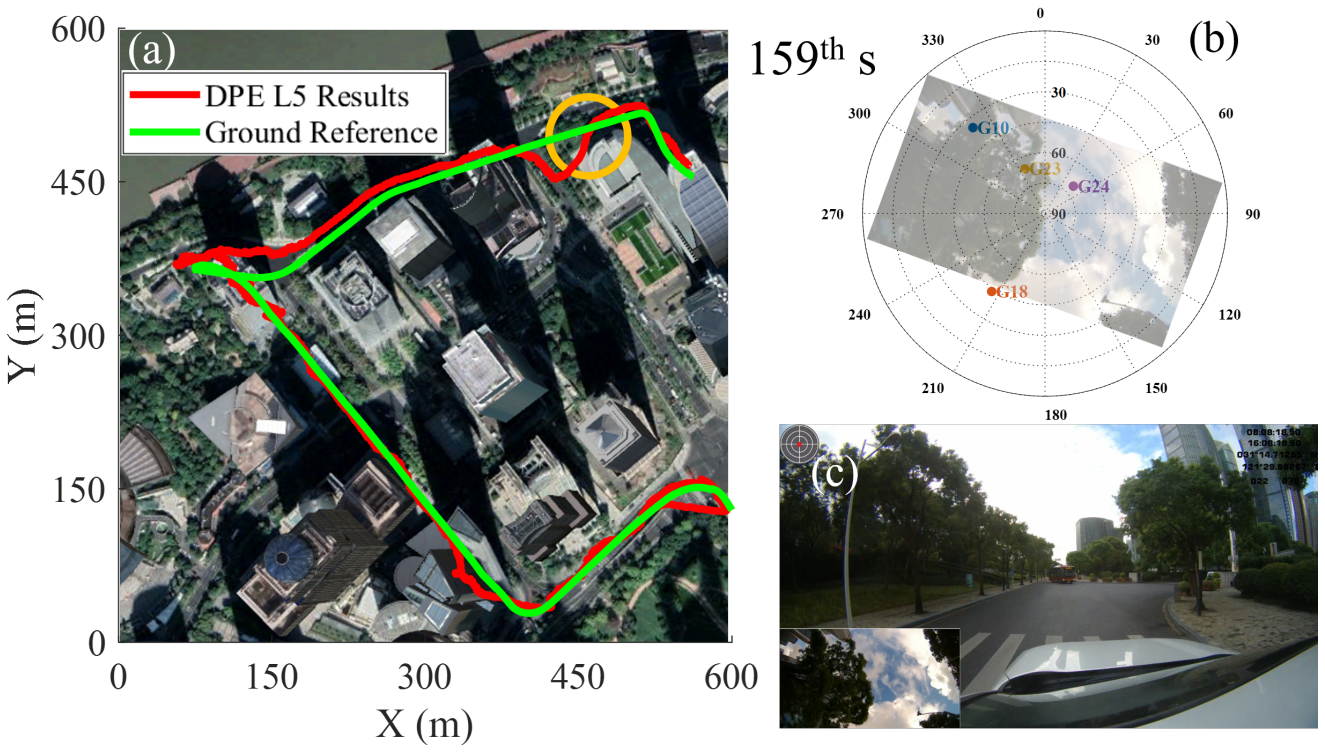

Fig. 14: The DPE PVT solutions versus receiver trajectory (a), sky plot (b), and the road map at 159s (c) during the drive test conducted in the Lujiazui area of Shanghai on September 3, 2022.

Fig. 14 presents the street view of the experimental data collection and the DPE positioning results using GPS L5 signals, alongside the ground reference derived from Xsens multi-sensor integrated navigation solutions [20]. It is evident that the DPE positioning results exhibit strong fluctuation relative to the ground reference, particularly around the 159th second, as highlighted in Fig. 14(a), where the DPE positioning error significantly increases due to multipath effects. Further analysis of the satellite sky plot and signal collection environment, shown in Figs. 14(b) and (c), reveals a complex urban scenario with the existence of both foliage and high building. Signals from PRN#10 and PRN#23, which pass through foliage, may suffer from signal scattering and short multipath effects, whereas the signal from PRN#18 is likely affected by the reflections from surrounding building, while PRN#24, nearly at the zenith, provides a relatively ideal signal reception condition.

Fig. 15 presents the CAF and PVT results at the 159th second, with Fig. 15(a)-(c) depicting the position space and Fig. 15(d)-(f) illustrating the velocity space. Focusing on the position space first, Fig. 15(a) shows the CAF result for code delay, with correlation peaks plotted relative to PRN#24, which has the best signal quality.

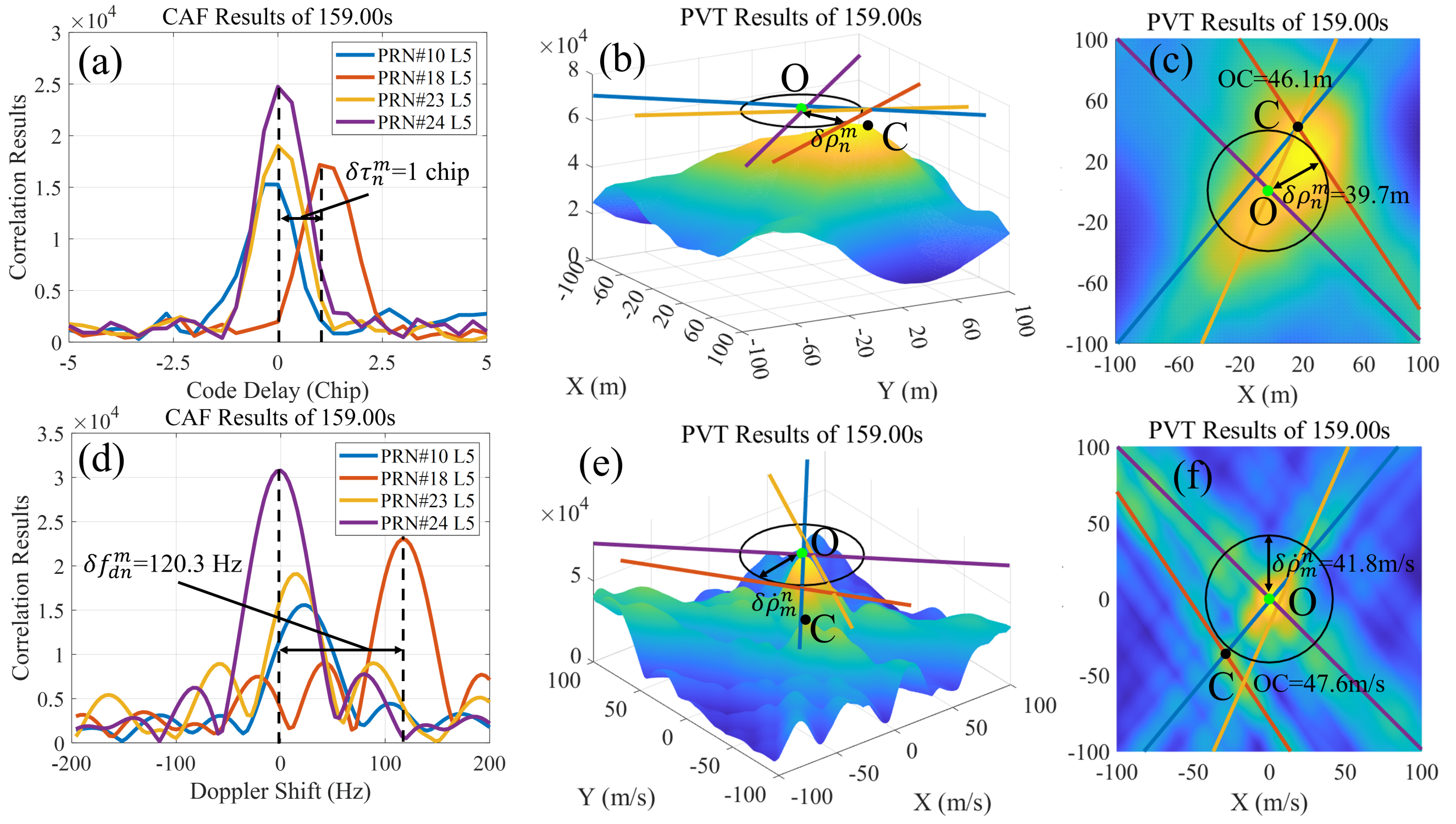

Fig. 15: The geometrical projection from CAF to PVT estimates in DPE receiver based on real GPS L5 data collected at 159s, shows a multipath code delay of 1 chip (a) and Doppler shift of 120Hz (b) are observed in CAF, which leads to a range and range rate bias at 39.7m and 41.8m/s, and the final PVT errors at 46.1m and 47.6m/s in 3D (b)(e) and 2D views (c)(f), respectively.

It is apparent that the signal from PRN#18 is a pure NLOS signal with an $\delta\tau_n^m$=1 chip multipath delay relative to the LOS center, while PRN#10 and PRN#23 exhibit slight offsets from the center due to scattering effects from foliage. Fig. 15(b) and (c) display the mapping of CAF to PVT and a 2D top-view projection of the PVT results, respectively. The theoretical correlation peak center lines align with the correlation results, and their intersections match the peak points of the correlation value energy superposition. The satellite circle centered at the true position O is tangent to the NLOS correlation peak center lines, validating the multipath geometry and SCMB model proposed in this paper using actual data. Moreover, the correlation values and PVT search area reveal that the multipath delay $\delta\tau_n^m$ mapped from CAF to a range bias of $\delta\rho_n^m$=39.7 m, while the positioning error $\delta r$, determined by the intersection point C, is away from the truth at the distance of 46.1 m, therefore, the positioning error $\delta r$=|OC|=46.1 m.

The velocity estimation errors shown in Fig. 15(d)-(f) are analyzed with PRN#24, the satellite with the best quality, as the reference. Fig. 15(d) reveals a distinct NLOS Doppler shift $\delta f_{dn}^m$=120.3 Hz for PRN#18, while PRN#10 and PRN#23 display similar slight offsets from the LOS center, likely due to scattering effects. Figs. 15(e) and (f) illustrate the mapping of CAF to PVT in velocity space, both in 3D and 2D views. Given that the INS-derived reference velocity might differ from the GNSS-calculated velocity, we use signal energy superposition as the reference. As shown in Fig. 15(f), the theoretical correlation peak center lines match the correlation results, validating the multipath geometry and SCMB model in the velocity space. The Doppler shift $\delta f_{dn}^m$ maps to a range rate bias $\delta\dot{\rho}$=41.8 m/s, and the velocity estimation error $\delta\dot{r}$, determined by the distance |OC| from the intersection point C to the circle center, is $\delta\dot{r}$=47.6 m/s.

TABLE VI: Theoretical and realistic error estimation from PRN#18

| $\delta\tau_n^m$ | $\delta\rho_n^m$ Theoretical/Real | $\delta r$ Theoretical/Real |
|---|---|---|
| 1 chip | 39.9/39.7 m | 47.2/46.1 m |
| $\delta f_{dn}^m$ | $\delta\dot{\rho}_n^m$ Theoretical/Real | $\delta\dot{r}$ Theoretical/Real |
| 120.3 Hz | 41.8/41.8 m/s | 49.5/47.6 m/s |

In addition to the experimental error estimates, theoretical errors can be derived using the geometric error propagation theory defined in (28), (29), (34), and (35). Given the multipath delay $\delta\tau_n^m$ and Doppler shift $\delta f_{dn}^m$ observed from PRN#18's NLOS CAF in Fig. 15(a) and (d), along with its elevation angle $\phi^m = 42.8°$ from Table II, the theoretical range bias $\delta\rho_n^m$ and range rate bias $\delta\dot{\rho}_n^m$ are calculated to be 39.9m and 41.8m/s, respectively, according to (28) and (29).

Using the intersection point C from PRN#18 and PRN#23 as the candidate PVT solution, and the azimuth angle difference $\Delta\theta = 122.3°$ from Table II, the theoretical PVT errors are computed as $\delta r = 47$m and $\delta\dot{r} = 47.6$m/s by substituting $\delta\rho_n^m$, $\delta\dot{\rho}_n^m$ and $\Delta\theta$ into (34) and (35).

Table VI presents both the theoretical and experimental estimations, demonstrating close alignment between the two sets of results and thereby validating the proposed multipath error propagation theory from a practical perspective.

## V. Conclusion

This paper systematically investigates the geometric characterization of multipath error propagation in DPE theory, and successfully quantifies the effects of the geometric distribution in error projection and propagation. Firstly, the geometric error mapping process from multipath delay and Doppler shift in CAF, to the range and range rate bias is examined. By modeling the NLOS correlation peak center line as a function of azimuth and elevation angles, a geometric expression for the CAF projection is derived. The findings reveal that the projection from CAF bias to the range and range rate bias is solely related to elevation but independent of azimuth. Additionally, with a constant CAF bias, the PVT bias increases with elevation, offering guidance for satellite selection in DPE under multipath conditions.

Secondly, the geometric propagation from the range or range rate errors to the PVT errors is explored through the theoretical derivation. The theoretical prediction shows that this mapping step is solely dependent on azimuth instead of the elevation. For performance analysis and verification, a satellite circle multipath bias (SCMB) model is proposed, where the receiver truth serves as center, and the range or range rate bias serves as radius, with the CAF projection lines tangent to the concentric circles. This model provides a intuitive approach for PVT error analysis.

Thirdly, building upon the SCMB model, the paper analyzes PVT error bounds under diverse multipath conditions using algebraic geometry. The theoretical analysis indicates that the minimum positioning error for two satellites affected by multipath depends on the satellite with the larger PVT bias, while the maximum error can approach infinity. When more than two satellites are involved, the minimum positioning error depends on the satellite with the second smallest PVT bias.

The paper offers a preliminary exploration of multipath error propagation in DPE framework, highlighting the potential of geometric theory for analyzing these errors in future applications. However, discrepancies between theoretical predictions and field test results were noted, likely due to insufficient consideration of signal energy in the error propagation analysis. Future research will aim to advance the comprehensive development on DPE error propagation, particularly in a variety of challenging environments.

## VI. Acknowledgments

This research was jointly supported by the National Science Foundation of China (No. 62003211) and the National Science Foundation of Shanghai (No.22ZR1434500).